\documentclass[12pt]{article}
\usepackage{amsfonts}
\usepackage{amssymb}

\def\hybrid{\topmargin -20pt    \oddsidemargin 0pt
        \headheight 0pt \headsep 0pt
        \textwidth 6.25in       
        \textheight 9.25in       
        \marginparwidth .875in
        \parskip 5pt plus 1pt   \jot = 1.5ex}

\hybrid

\def\baselinestretch{1.2}

\catcode`\@=11

\def\marginnote#1{}
\newcount\hour
\newcount\minute
\newtoks\amorpm
\hour=\time\divide\hour by60
\minute=\time{\multiply\hour by60 \global\advance\minute by-\hour}
\edef\standardtime{{\ifnum\hour<12 \global\amorpm={am}%
        \else\global\amorpm={pm}\advance\hour by-12 \fi
        \ifnum\hour=0 \hour=12 \fi
        \number\hour:\ifnum\minute<10 0\fi\number\minute\the\amorpm}}
\edef\militarytime{\number\hour:\ifnum\minute<10 0\fi\number\minute}

\def\draftlabel#1{{\@bsphack\if@filesw {\let\thepage\relax
   \xdef\@gtempa{\write\@auxout{\string
      \newlabel{#1}{{\@currentlabel}{\thepage}}}}}\@gtempa
   \if@nobreak \ifvmode\nobreak\fi\fi\fi\@esphack}
        \gdef\@eqnlabel{#1}}
\def\@eqnlabel{}
\def\@vacuum{}
\def\draftmarginnote#1{\marginpar{\raggedright\scriptsize\tt#1}}

\def\draft{\oddsidemargin -.5truein
        \def\@oddfoot{\sl preliminary draft \hfil
        \rm\thepage\hfil\sl\today\quad\militarytime}
        \let\@evenfoot\@oddfoot \overfullrule 3pt
        \let\label=\draftlabel
        \let\marginnote=\draftmarginnote
   \def\@eqnnum{(\theequation)\rlap{\kern\marginparsep\tt\@eqnlabel}%
\global\let\@eqnlabel\@vacuum}  }

\def\preprint{\twocolumn\sloppy\flushbottom\parindent 2em
        \leftmargini 2em\leftmarginv .5em\leftmarginvi .5em
        \oddsidemargin -.5in    \evensidemargin -.5in
        \columnsep .4in \footheight 0pt
        \textwidth 10.in        \topmargin  -.4in
        \headheight 12pt \topskip .4in
        \textheight 6.9in \footskip 0pt
        \def\@oddhead{\thepage\hfil\addtocounter{page}{1}\thepage}
        \let\@evenhead\@oddhead \def\@oddfoot{} \def\@evenfoot{} }

\def\numberbysection{\@addtoreset{equation}{section}
        \def\theequation{\thesection.\arabic{equation}}}

\def\underline#1{\relax\ifmmode\@@underline#1\else
        $\@@underline{\hbox{#1}}$\relax\fi}

\def\titlepage{\@restonecolfalse\if@twocolumn\@restonecoltrue\onecolumn
     \else \newpage \fi \thispagestyle{empty}\c@page\z@
        \def\thefootnote{\fnsymbol{footnote}} }

\def\endtitlepage{\if@restonecol\twocolumn \else \newpage \fi
        \def\thefootnote{\arabic{footnote}}
        \setcounter{footnote}{0}}  

\catcode`@=12
\relax

\def\figcap{\section*{Figure Captions\markboth
        {FIGURECAPTIONS}{FIGURECAPTIONS}}\list
        {Figure \arabic{enumi}:\hfill}{\settowidth\labelwidth{Figure
999:}
        \leftmargin\labelwidth
        \advance\leftmargin\labelsep\usecounter{enumi}}}
 \relax
\def\tablecap{\section*{Table Captions\markboth
        {TABLECAPTIONS}{TABLECAPTIONS}}\list
        {Table \arabic{enumi}:\hfill}{\settowidth\labelwidth{Table
999:}
        \leftmargin\labelwidth
        \advance\leftmargin\labelsep\usecounter{enumi}}}
 \relax
\def\reflist{\section*{References\markboth
        {REFLIST}{REFLIST}}\list
        {[\arabic{enumi}]\hfill}{\settowidth\labelwidth{[999]}
        \leftmargin\labelwidth
        \advance\leftmargin\labelsep\usecounter{enumi}}}
 \relax
\makeatletter
\newcounter{pubctr}
\def\publist{\@ifnextchar[{\@publist}{\@@publist}}
\def\@publist[#1]{\list
        {[\arabic{pubctr}]\hfill}{\settowidth\labelwidth{[999]}
        \leftmargin\labelwidth
        \advance\leftmargin\labelsep
        \@nmbrlisttrue\def\@listctr{pubctr}
        \setcounter{pubctr}{#1}\addtocounter{pubctr}{-1}}}
\def\@@publist{\list
        {[\arabic{pubctr}]\hfill}{\settowidth\labelwidth{[999]}
        \leftmargin\labelwidth
        \advance\leftmargin\labelsep
        \@nmbrlisttrue\def\@listctr{pubctr}}}
 \relax
\makeatother
\newskip\humongous \humongous=0pt plus 1000pt minus 1000pt

\newif\ifdtup

\relax

\def\be{\begin{equation}}
\def\ee{\end{equation}}
\def\ba{\begin{eqnarray}}
\def\ea{\end{eqnarray}}

\def\del{\partial}

\def\r{\rho}
\def\a{\alpha}

\def\b{\beta}

\def\G{\Gamma}
\def\d{\delta}

\def\e{\epsilon}

\def\p{\pi}
\def\P{\Pi}

\def\th{\theta}
\def\Th{\Theta}
\def\m{\mu}
\def\n{\nu}
\def\om{\omega}
\def\Om{\Omega}
\def\l{\lambda}
\def\L{\Lambda}
\def\s{\sigma}
\def\S{\Sigma}

\def\cR{{\cal R}}
\def\cA{{\cal A}}

\def\bs{\bigskip}
\def\no{\noindent}

\def\qq{\qquad}

\def\IR{\relax{\rm I\kern-.18em R}}

\def \ha {{1\over 2}}

\def \ov {\over}

\def\const{{\rm const.}}

\def\IR{\relax{\rm I\kern-.18em R}}
\def\inv{^{\raise.15ex\hbox{${\scriptscriptstyle -}$}\kern-.05em 1}}

\def\cR{{\cal R}}
\def\cA{{\cal A}}

\begin{document}

\renewcommand{\theequation}{\thesection.\arabic{equation}}

\newcommand{\beq}{\begin{equation}}
\newcommand{\eeq}[1]{\label{#1}\end{equation}}
\newcommand{\ber}{\begin{eqnarray}}
\newcommand{\eer}[1]{\label{#1}\end{eqnarray}}
\newcommand{\eqn}[1]{(\ref{#1})}
\begin{titlepage}
\begin{center}

\hfill CPHT-RR044.0705\\
\vskip -.1 cm
\hfill hep--th/0507134\\

\vskip  .7 in

{\large \bf Fundamental branes and shock waves}

\vskip 0.4in

{\bf Konstadinos Sfetsos}

\vskip .1in

Department of Engineering Sciences, University of Patras\\
26110 Patras, Greece\\
\vskip .1in
and \\
\vskip .1in
Centre de Physique Th\'eorique, \'Ecole Polytechnique\\
91128 Palaiseau, France\\

\end{center}

\vskip .3in

\centerline{\bf Abstract}

\no
We construct supersymmetric brane solutions in string and M-theory with
moduli parameters that depend arbitrarily on the light-cone time.
Our investigation aims in understanding time dependent phenomena in
gauge theories at strong coupling within the gauge/gravity correspondence.
For that reason we use, as
a basic ingredient, multicenter supergravity solutions which model
the Coulomb branch of the corresponding strongly coupled gauge theories.
We introduce the notion of shape invariant motions and show that in a
particular limit involving pulse-type motions of finite energy,
the solutions represent
gravitational shock waves moving on the brane background geometry.
We apply the general formalism for
D3-branes distributed on a disc and on a sphere as well as for NS5-branes
distributed on a ring, all with time varying radii.
We examine the problem of open strings attached
on moving branes and suggest a mechanism which may be responsible
for giving rise at a macroscopic level to gravitational shock waves.

\noindent

\vskip .5in

\noindent
July 2005\\
\end{titlepage}
\vfill
\eject

\def\baselinestretch{1.2}
\baselineskip 17.5 pt
\noindent


\section{Introduction}
\setcounter{equation}{0}
\renewcommand{\theequation}{\thesection.\arabic{equation}}

The gauge/gravity correspondence \cite{Maldacena}
has been used over the last years mainly in
order to study gauge theories at strong coupling.
Going beyond the original
proposal on the holographic relation between $N=4$ SYM and string
theory on $AdS_5\times S^5$ required the introduction of moduli parameters
that break conformality, as well as part, if not all, of the supersymmetry and
global symmetries.
A natural question that will be the main focus of this paper is whether or not
we can promote these moduli parameters to functions of time since
understanding time dependent phenomena in physics is of
immense importance. Given that the vast majority of tests and
predictions within the gauge/gravity correspondence concerned supersymmetric
field theories and their supergravity duals, we would like, at least in a first
step, to introduce time dependence in a supersymmetric manner in addition to
satisfying the field equations.
This is a strong condition and as a result the time dependence comes
in the form of the light-cone time, but it is otherwise arbitrary.
In this paper we will be concerned with the construction of supergravity
solutions with light-cone dependent moduli parameters.
Among the various supergravity solutions with field theory duals
we could have chosen to start our investigations,
perhaps the most attractive for our purposes are those that deviate the least
from the conformal supersymmetric case.
To be specific we will use multicenter supergravity
solutions representing the gravitational field of a large collection of
fundamental branes in string and M-theory with unwrapped all the
worldvolume directions. The constant moduli parameters
that will be promoted to functions of the light-cone time are nothing but the
centers of these branes which
from the gauge theoretical point o view represent
vacuum expectation values of scalar fields \cite{trivedi,sfet1,klewi}.

This paper is organized as follows: In section 2 we present the main idea
and results that apply generically to all fundamental branes with
centers depending on the light-cone time.
In sections 3 and 4 we derive and present in detail
the light-cone time dependent supergravity solutions for
all fundamental branes of M- and string theory. The emphasis is given on
the preservation of a fraction of supersymmetry compared with the
generic static case.
In section 5 we consider an important subclass of center motions
in which the angles between the vectors, defining them in the transverse to
the brane space, remain constant. Therefore, in these cases the shape of the
distribution of the centers remains invariant.
We also consider the limit of sudden
changes in which the supergravity solution reduces to that of a shock wave
propagating on the background multicenter brane
solution and compute its profile in general. In section 6 we explicitly
construct the solutions for the cases of D3-branes uniformly and continuously
distributed on a disc and on a 3-sphere of varying radii.
In addition we construct the
shock waves on the maximally supersymmetric spaces $AdS_5\times S^5$,
$AdS_{4,7}\times S^{7,4}$ of the ten-dimensional type-IIB and
eleven-dimensional sypergravities, respectively
and consider scattering amplitudes in the propagation of scalar fields.
In section 7 we consider the case of NS5-branes
on a circle of varying radii.
We explicitly solve the eigenvalue problem
for the massless field spectrum and use the result for the computation of
the exact scattering shock wave amplitudes.
In section 8 we consider the problem of the
first quantized open string with one end on a moving brane and compute exactly
its unitary evolution in time for arbitrary brane motion.
Finally, we present our conclusions and directions for future work in
section 9.
The paper is supplemented with an Appendix where we have collected the
expressions for the spin connection and the Ricci tensor for
the general form of the supergravity backgrounds we consider.

\section{The general construction}
\setcounter{equation}{0}
\renewcommand{\theequation}{\thesection.\arabic{equation}}

Before we turn into the description of our brane solutions in detail, we
present in this section some general aspects and results of the construction.
The supersymmetric branes of M-theory and string theory with
unwrapped all the world-volume directions, correspond to
supergravity solutions whose metric can be cast in the following
general form (for reviews, see, for instance, \cite{StelleTow})
\ba
&& ds^2 = H^\a (-dt^2+dy_1^2+\cdots +dy_{p}^2 )+
H^{1+\a} dx^i dx^i\ ,
\nonumber\\
&&
i=1,2,\dots , d\ , \qq H=H(x)\ .
\label{me1}
\ea
The dimensionality of the space-time is $D=p+1+d$ and $H$ is a harmonic
function in the transverse space $\IR^d$, i.e.
$\del^2 H=0$. The numerical parameter $\a$ depends on the
particular brane and the metric is supported by a non-trivial flux
and possibly a dilaton
so that the classical supergravity equations of motion are
satisfied and a fraction of the maximal supersymmetry is preserved.
We may introduce a wave by writing $-dt^2 +dy_1^2=2 dudv$,
denoting the rest of the directions along the brane by
$y^\a$, $\a=2,\dots , p$ and generalizing the ansatz \eqn{me1}
to\footnote{There should be no confusion between the numerical
parameter $\a$ and the
space-time index denoted by the same symbol.}
\ba
&&  ds^2 = H^\a \Big[dy^\a dy^\a+ 2 du dv + F(x,u) du^2 + 2
V_i(x,u) dx^i du\Big] + H^{1+\a} dx^i dx^i\ ,
\nonumber\\
&& i=1,2,\dots , d\ ,\qq \a=2,\dots, p\ , \qq
H=H(x,u)\ ,\quad V_i=V_i(x,u)\ .
\label{js45}
\ea
Hence, the various functions are allowed to depend on the transverse space
coordinates $x^i$ as well as on the light-cone time $u$.
The presence of the wave breaks the Lorentz symmetry along the brane to its
$SO(p-1)$ rotational subgroup times an $\IR$ factor
representing shifts of $v$.
We aim at constructing supersymmetric solutions by supporting the metric
\eqn{js45} with appropriate tensor fields.
The gravitino and dilatino Killing spinor equations are of the general form
\ba
&& \del_\m\e + {1\ov 4}\om^{ab}_\m \G_{ab}\e +(\cdots )\e= 0\ ,
\nonumber\\
&& \G^\m\del_\m \Phi\e + (\cdots)\e = 0 \ ,
\label{Kiill}
\ea
where the extra terms depend on products of the tensor fields and
$\G$-matrices. In these computations we will use the general frame
\ba
&&  e^+ = H^{\a/2} du\ ,
\qq  e^- =H^{\a/2}\left( dv + V_i dx^i +{F\ov 2}du \right)\ ,
\nonumber\\
&&
 e^i= H^{(1+\a)/2} dx^i\ , \qq e^{\a}= H^{\a/2} dy^\a \ .
\label{frra}
\ea
The spin connection necessary for solving the Killing spinor equations as
well as the Ricci tensor are computed in the Appendix.
We will find that it is indeed possible to construct such solutions
and the amount of supersymmetry to
be preserved depends on the dimensionality of the brane world-volume.
Typically, the solution in the absence of the
wave preserves half of the maximal supersymmetry.
If the world-volume is a least ($2+1$)-dimensional, the presence
of the wave requires an extra projection which is provided by
\be
\G^+\e=0\ .
\label{pprr}
\ee
This conclusion applies to M2- and M5-branes, the NS5-branes, as
well as all Dp-branes with $p\geq 2$. Hence in these cases our
configurations will preserve ${1\ov 4}$ of the maximum supersymmetry.
For the fundamental string
NS1 and the D1-brane the amount of supersymmetry is
the same as in the case of no wave, i.e. $\ha$ of the maximum,
and \eqn{pprr} provides
the only projection since $\G^{01}\e=\e$ is equivalent to \eqn{pprr}.
The general form of the Killing spinor that satisfies \eqn{Kiill}
is essentially dictated by the
supersymmetry algebra (as noted in a similar context in \cite{kumar}) and reads
\be
\e=H^{\a/4}\e_0\ ,
\ee
where $\e_0$ is a constant spinor subject to the same projections as $\e$.
The Killing spinor eq. determines
the form of the tensor fields corresponding to the branes as well as in the
case of string branes the dilaton field. However, it is not sufficient to
completely determine the unknown functions $V_i$ and $F$ in the background
metric \eqn{js45}. The reason is essentially that for
Lorenzian backgrounds the integrability condition for the Killing spinor
equation does not automatically imply the second order supergravity equations
of motion. Hence, we have to also employ the Einstein field equations
$R_{\m\n}=\cdots$. It turns out that, for all fundamental branes of M- and
string theory there is a general result
\be
\del^2 H = \del^2 F=\del^2 V_i=0 \ ,
\label{hdv1}
\ee
that is, all functions entering into the expression for the metric
\eqn{js45} are harmonic in the $d$-dimensional transverse space to the branes.
There is an additional condition linking together $H$ and $V_i$, namely
\be
{\dot H}= \del_i V_i \ ,
\label{hdv}
\ee
where the dot represents the derivative with
respect to $u$.\footnote{It turns out that this is the most convenient gauge choice
to present our results in a clear way.
The conditions \eqn{hdv1} have been written after using \eqn{hdv}. See also
the transformation \eqn{jhhh1}-\eqn{bfb3} below and the related
comment at the end of this section.}
Therefore the general asymptotically flat
solution representing the gravitational field
of $N$ branes has
\ba
&& H(x,u)=1+{\cA}_{\rm sc}\sum_{a=1}^N {1\ov |{\bf x}-{\bf x}_a(u)|^{d-2}}\ ,
\nonumber\\
&& V_i(x,u) = -{\cA}_{\rm sc}
\sum_{a=1}^N {\dot x^i_a(u) \ov |{\bf x}-{\bf x}_a(u)|^{d-2}}\ ,
\ea
where the $d$-dimensional vectors ${\bf x}_a$, $a=1,2,\dots , N$,
correspond to the
location of the branes in the transverse space.
Hence we see that compared with the static case the main difference is that
the moduli vectors representing the constant positions of the branes
have been promoted to arbitrary functions ${\bf x}_a(u)$ of the light-cone
time.\footnote{Essentially the end result reminds of the explicit
construction of manifolds with Lorentzian holonomy in various dimensions,
where also the promotion of constant moduli parameters in Euclidean holonomy
manifolds into arbitrary functions of the
light-cone time, played an essential r\^ole \cite{hersfe}.}
The constant ${\cA}_{\rm sc}$, with
units of $\rm (length)^{d-2}$, depends
on the Planck scale for M-branes, the string scale and the string coupling
constant for the branes of string theory, as well as on numerical factors.
This constant does not depend on the fact that we have
introduced lightcone dependence to the moduli parameters and has the same
value as in the static case (see, for instance, \cite{malth}).
From now on we scale for convenience the constant ${\cA}_{\rm sc}$
to unity.
Also note that, we have not included a constant part in the $V_i$'s
since, unlike in $H$, this can be absorbed by a shift of the coordinate $v$.
Finally, we note that the functions $F$ and $V_i$ in the general metric
\eqn{js45} can all be set to zero by a suitable coordinate
transformation (as in the case of backgrounds with a covariantly
constant null Killing vector \cite{Bri} and also in \cite{Tseytlin,Horowitz:1994rf}.
Indeed, let
\be
x^i =f^i(u,x')\ ,\qq v=v'+h(u,x')\ .
\label{jhhh1}
\ee
Then for the new coordinates $(u,v',x'^i)$ we obtain the same metric as
in \eqn{js45} but with
\ba
F'(u,x') & = & F+2 \dot h +\dot f^i \dot f^i +2 V_i\dot f^i \ ,
\nonumber\\
 V'_i(u,x') & = & \del_i h + V_j \del_i f^j + \del_i f^j \dot f^j \ .
\label{jhhh2}
\ea
Setting $F'=V_i'=0$ in \eqn{jhhh2},
gives a first order system for the unknown functions
$h$ and $f^i$ which in principle can be solved. However,
there is a price to pay. Namely,
the flat transverse space metric $dx^idx^i$ is replaced by the curved one
\be
g'_{ij}(u,x') =  \del_i f^k \del_j f^k \ .
\label{bfb3}
\ee
This last, rather undesirable feature, makes the form of the metric \eqn{js45}
preferable for the general discussion.
Finally, note that for $f_i=x_i$ in \eqn{jhhh1}-\eqn{bfb3} we have a gauge-like
transformation with parameter $h$ and the transverse space metric remains flat.
After choosing \eqn{hdv}, only $h$'s satisfying $\del^2h=0$ preserve this choice
and the conditions \eqn{hdv1}. This remaining freedom still allows to set the function
$F=0$, but we will not do so since it will be necessary for presenting
specific classes of examples below. Another reason, not of direct concern to us,
is that in the special case that $\del/\del u$ is also a Killing vector,
the metric \eqn{js45} cannot be written in a way that is manifestly independent of $u$
and simultaneously general, unless $F\neq 0$ (for an analogous discussion see
also \cite{Horowitz:1994rf}).

\section{M-theory branes}
\setcounter{equation}{0}
\renewcommand{\theequation}{\thesection.\arabic{equation}}

In the context of eleven-dimensional supergravity \cite{Cremmer}
the geometry is supplemented by a 4-form field
strength $F_{\m\n\r\s}$.
The Killing spinor equation arising by setting to zero the gravitino
superymmetry variation is
\be
\del_\m\e + {1\ov 4}\om^{ab}_\m \G_{ab}\e -{1\ov 288}
\left(F_{\n\r\l\s}\G^{\n\r\l\s}{}_\m -8 F_{\m\n\r\l}\G^{\n\r\l}\right)\e=0\ .
\label{kiil11}
\ee
In addition, the Einstein equations of motion
\be
R_{\m\n}={1\ov 12} \left((F^2)_{\m\n} -{1\ov 12} g_{\m\n} F^2\right)\ ,
\ee
together with that for the 4-form field strength should be satisfied.

\subsection{M2-branes}

In this case the metric is given by \eqn{me1} with
$\a=-2/3$ and $d=8$, $p=2$, so that $D=11$. In the static limit
it describes the multicenter
generalization of the M2-brane solution of \cite{M2}. Working out the
details of the Killing spinor eq. \eqn{kiil11}
we find that the usual projection
\be
\G^{+-2}\e = \e\ ,
\ee
where all indices are along the brane, should be supplemented
with \eqn{pprr}. This means that $\G^2\e=\e$.
Also we get
\be
F_{+-2i}=-H^{-7/6} \del_i H\ , \qq F_{+2ij}=H^{-2/3}\del_{[i} V_{j]}\ ,
\ee
where all the indices refer to tangent space.

\no
We note that all other M- and string theory brane solutions
considered below follow from the M2-brane solution
after performing a series of operations involving dimensional reductions,
a smearing along a direction transverse to the brane and U-dualities.
This step by step procedure is well known and rather straightforward
for brane solutions with
a single center, but it can also be generalized in the multicenter
case.
Nevertheless, we preferred to be analytic and pedagogical
in our approach so that
all different cases will be obtained in an independent way.

\subsection{M5-branes}

In this case the metric is given by \eqn{me1} with
$\a=-1/3$ and $d=5$, $p=5$, so that $D=11$.
In the static limit it describes the multicenter
generalization of the M2-brane solution of \cite{M5}.
Working out the
details of the Killing spinor eq. \eqn{kiil11}
we find that the usual projection (all indices are transverse to the
brane)
\be
\G^{12345}\e =- \e\ ,
\ee
when supplemented with \eqn{pprr} gives rise to
\be
F_{ijkl}=H^{-4/3} \e_{ijklm}\del_m H\ , \qq
F_{+ijk}=-\ha H^{-5/6}\e_{ijklm}\del_{[l} V_{m]}\ ,
\ee
where all the indices refer to tangent space.

\section{String theory branes}
\setcounter{equation}{0}
\renewcommand{\theequation}{\thesection.\arabic{equation}}

Our discussion concerning string theory branes will be in the context of
type-IIA \cite{Huq, Giani, howe2} or IIB \cite{Schwarz,Howe} supergravities
(for a pedagogical treatment in relation also to
T-duality in the presence of RR fields, see also \cite{ortin}).

\subsection{Dp-branes ($p\neq 3$)}

In this case the metric is given by \eqn{me1} with
$\a=-1/2$ and $d=9-p$, so that $D=10$. In the static limit
it describes the multicenter
generalization of the $Dp$-brane solutions of \cite{Horowitz,Polchinski}.
The geometry is supplemented by a ($p+2$)-form field
strength $F_{p+2}$ in the RR-sector.
The Killing spinor equations arising by setting to zero the gravitino
and dilatino superymmetry variations in
type-IIA or IIB supergravity according to weather $p$ is even or odd,
respectively, are (we perform the computation in the string frame)
\cite{Behrndt}
\ba
&& \del_\m\e + {1\ov 4}\om^{ab}_\m \G_{ab}\e
- {(-i)^{p+1}\ov 8 (p+2)!} e^{\Phi} F_{p+2}\cdot \G\ \G_\m\e_{(p)}=0\ ,
\nonumber\\
&&\G^\m\del_\m\Phi \e + i^{p+1}{3-p\ov 4(p+2)!} e^\Phi F_{p+2}\cdot \G \e_{(p)}=0\ ,
\label{ssuu}
\ea
where
\be
F_{p+2}\cdot \G= F_{\m_1\cdots \m_{p+2}} \G^{\m_1\cdots \m_{p+2}}\ ,
\ee
and
\be
\e_{(1,5)}=i \e^*\ ,\qq \e_{(2,6)}=-\G_{11}\e \ ,\qq \e_{(-1,3,7)}=i \e\ ,\qq
\e_{(0,4,8)}=\e\ .
\label{jhew}
\ee
In addition, the Einstein equations of motion
\be
R_{\m\n}+ 2D_{\m}D_{\n}\Phi =  {e^{2\Phi}\ov 2 (p+1)!}
\Big((F_{p+2}^2)_{\m\n}-{1\ov 2 (p+2)} g_{\m\n}F_{p+2}^2\Big) \ ,
\label{dh1}
\ee
as well as that for the dilaton
\be
R+4 \left[D^2\Phi -(\del\Phi)^2\right] =0 \ ,
\label{sppl}
\ee
and the fluxes should be satisfied.
Here we are interested in the cases with $p\neq 3$.
For D3-branes there are some slight changes due to the self-dual
5-form field strength, which will be taken into account below.
We find the usual projection
\be
\G^{+-2\cdots p}\e_{(p)}= i^{p+1}\e\
\ee
and in addition \eqn{pprr}, whereas for the $(p+2)$-form we obtain
\be
F_{+-2\cdots pi}=H^{p/4}\del_iH^{-1}\ ,\qq
F_{+2\cdots p ij}= (-1)^p H^{(p-6)/4} \del_{[i}V_{j]}\ ,
\ee
where all indices refer to the tangent space.
Consistency requires also a nontrivial dilaton given by
\be
e^{-2 \Phi}=H^{p-3\ov 2}\ .
\ee

\subsection{D3-branes}

In this case we have to satisfy that
\be
F_{5}=\pm A +*A\ ,
\ee
for some 5-form field strength $A$, so that $F_5$ is selfdual (anti-selfdual).
The supersymmetry variation for the gravitino is
\ba
\del_\m\e + {1\ov 4}\om^{ab}_\m \G_{ab}\e
+ {i\ov 480}  F_{\m_1\cdots \m_5}\G^{\m_1\cdots \m_5} \G_\m\e=0\ .
\label{s5u}
\ea
In addition, we should satisfy the Einstein equations of motion
\be
R_{\m\n}={1\ov 6} (F^2)_{\m\n}={1\ov 3}\left( (A^2)_{\m\n}-{1\ov 10} g_{\m\n}
A^2\right)\ .
\label{lkwe}
\ee
Using that
\be
F_{\m_1\cdots \m_5}\G^{\m_1\cdots \m_5}=\pm 2
A_{\m_1\cdots \m_5}\G^{\m_1\cdots \m_5}\ .
\label{jde}
\ee
and taking into account \eqn{jhew}, the Killing spinor eq.\eqn{ssuu} gives
the projection
\be
i \G^{+-12}\e=\e\ ,
\ee
in addition to \eqn{pprr}, so that $i \G^{12}\e=\e$. Also
\be
A_{+-12i}=\pm {1\ov 4} H^{-5/4} \del_iH\ ,\qq
A_{+12ij}=\pm {1\ov 4} H^{-3/4} \del_{[i} V_{j]}\ .
\ee
Note that the form of the Killing spinor eq. \eqn{s5u} when written for $A$
using \eqn{jde} as well as the field equation \eqn{lkwe} correspond to
\eqn{ssuu} and \eqn{dh1} (for $p=3$ and $\Phi=\const$)
after the identification $A=\mp F_{p+2}/4$.
Also, from \eqn{lkwe} the scalar curvature $R=0$ which from \eqn{sppl} shows
that it is consistent to set $\Phi=\const$.

\subsection{String theory NS-branes}
\renewcommand{\theequation}{\thesection.\arabic{equation}}

In this case the geometry is supplemented by a 3-form field
strength $H_{\m\n\r}$ and a dilaton field.
The Killing spinor equations arising by setting to zero the gravitino and
dilatino supersymmetry variations are
\ba
&& \del_\m\e + {1\ov 4}(\om^{ab}_\m -\ha H_\m{}^{ab})\G_{ab}\e= 0\ ,
\nonumber\\
&& \G^\m \del_\m\Phi \e -{1\ov 12} H_{\m\n\r}\G^{\m\n\r}\e=0\ .
\label{kns}
\ea
In addition, the equations of motion
\ba
&& R_{\m\n}-{1\ov 4} (H^2)_{\m\n}+2 D_\m D_\n\Phi = 0\ ,
\nonumber\\
&& D_\m \left(e^{-2 \Phi}H^\m{}_{\n\r}\right)=0\ ,
\ea
should be satisfied.

\subsubsection{NS1-branes}

In this case $\a=-1$ and $d=8$, $p=1$ so that $D=10$. Using the
Killing spinor eq. \eqn{kns}
we find that the only projection to be imposed is \eqn{pprr}. In addition
we determine the dilaton field and the antisymmetric tensor field strength as
\be
e^{-2 \Phi}=H\
\label{dj2}
\ee
and
\be
H_{+-i}=-H^{-1}\del_iH\ ,\qq H_{+ij}=-H^{-1/2} \del_{[i}V_{j]}\ ,
\ee
where all the indices refer to tangent space.
This can be derived by an antisymmetric tensor field with non-vanishing
components
\be
B_{uv}=H^{-1}\ ,\qq B_{ui}= H^{-1} V_i \ .
\ee
This result was found before in \cite{callan}.

\subsubsection{NS5-branes}

In this case $\a=0$ and $d=4$, $p=5$, so that $D=10$. From
the Killing spinor eq. \eqn{kns}
we find that the usual projection (all indices are transverse to the brane)
\be
\G^{1234}\e =- \e\ ,
\ee
together with \eqn{pprr}. Also we find the dilaton
\be
e^{2 \Phi}=H\
\label{dj3}
\ee
and the antisymmetric tensor field strength components
\be
H_{ijk}=H^{-3/2} \e_{ijkl}\del_l H\ , \qq
H_{+ij}=-\ha H^{-1}\e_{ijkl}\del_{[k} V_{l]}\ .
\ee
Again all the indices refer to tangent space and \eqn{hdv} provides
the necessary for its integrability condition.

\section{Shape invariant motions and shock waves}
\setcounter{equation}{0}
\renewcommand{\theequation}{\thesection.\arabic{equation}}

In order to proceed we have to specify the vectors ${\bf x}_a$ as
functions of the light-cone time. Recall that in the static
case an arbitrary distribution of these centers breaks completely the
rotational $SO(d-2)$ symmetry of $\IR^{d-2}$. We are mostly interested in cases
where a continuous or a discrete subgroup of the full rotational group
can be preserved by the center distribution. In the non-static case,
even if such a subgroup is at some moment preserved,
an arbitrary
motion of the centers will cause a further destruction of the symmetry.
More interesting are cases where this can be kept under control.
For instance, we may consider motions in which in the far past and future
at $u\to \mp\infty$, a
given symmetry subgroup $H_{\mp \infty}\in SO(d-2)$ is
preserved. Then, it is interesting to investigate
the type of phenomena that arise in the transition between the two.
In this section (and paper) we will be concerned with an even simpler type of
center motion in which all of of them
change in time as a result of a single overall function, that is
\be
x^i_a(u)=r_0(u) x^i_a\ ,
\ee
where the $x^i_a$'s on the right hand side are constant moduli.
Such a motion leaves invariant the angles between the defining
vectors of the centers ${\bf x}_a$ and therefore,
although their distribution changes in time, it does
so in a {\it shape invariant} way.
To better study the solution we perform the following change of variables
\be
x_i\to r_0(u) x_i \ ,\qq (u,v)\to (u,v)\ .
\ee
As a result in the metric \eqn{js45} the coefficient of the $du^2$ term
becomes
\be
F(u,x)=\dot r_0^2 {\bf x}^2 + {\dot r_0^2\ov r_0^{d-2}}
\sum_a {{\bf x}^2- 2 {\bf x}\cdot {\bf x}_a\ov |{\bf x}-{\bf x}_a|^{d-2}}\ ,
\ee
where we have assumed that initially $F=0$ (but $V_i\neq 0$). Similarly,
the coefficient of the $2 dudx^i$ term changes to
\be
V_i\to r_0 \dot r_0 x^i + {\dot r_0\ov r_0^{d-3}}
\sum_a {x^i-x^i_a\ov |{\bf x}-{\bf x}_a|^{d-2}}\ .
\ee
However, this is a total derivative
\be
V_i=-\del_i \L \ ,\qq \L= -\ha r_0\dot r_0 {\bf x}^2
+ {1\ov d-4} {\dot r_0\ov r_0^{d-3}}
\sum_a {1\ov |{\bf x}-{\bf x}_a|^{d-4}}\ ,
\ee
which can be eliminated by the coordinate shift $v\to v+ \L$.
This, of course affects the coefficient of the $du^2$ term.
After taken all these into account we find the end result for the metric
\be
 ds^2 = H^\a \Big[dy^\a dy^\a+
2 du dv + F(x,u) du^2 \Big] + r_0^2 H^{1+\a} dx^i dx^i\ ,
\label{jsner}
\ee
where
\be
H(u,x)=1+ {1\ov r_0^{d-2}}
\sum_{a=1}^N {1\ov |{\bf x}-{\bf x}_a|^{d-2}}\
\label{dllsh}
\ee
and
\be
F(u,x)=-r_0\ddot r_0 {\bf x}^2 -{\dot r_0^2\ov r_0^{d-2}}
\sum_a {{\bf x}_a^2\ov |{\bf x}-{\bf x}_a|^{d-2}}
+ { 2\ov d-4 }r_0^{2-d/2}
\del_u \left(\dot r_0\ov r_0^{d/2-1}\right)
\sum_a{1\ov |{\bf x}-{\bf x}_a|^{d-4}} \ .
\label{jhf4}
\ee
This geometry is supported by fluxes with the same non-vanishing
tangent space components as in the static case (since $V_i=0$).
In other words for shape invariant motions of the centers it turns out that
the matter flux fields supporting the metric for a
valid gravitational solution to exist, remain unchanged.
In the cases of NS1, NS5
and $Dp$-branes (with $p\neq 3$) there is also a dilaton field which
however, due to the fact that it depends solely on the harmonic function $H$
in \eqn{dllsh}, it remains $u$-dependent.

Let's consider the cases of $M2$, $M5$, $D3$ and $NS5$ branes, in the field
theory limit in which effectively the unity in \eqn{dllsh}
is ignored and as a consequence
also the first term in \eqn{jhf4}. In addition, let's perform the coordinate
transformation
\be
u\to \int^u du' [r_0(u')]^{4-d}\ , \qq
y^\a\to y^\a r_0^{2-d/2}\ ,\qq v\to v +(d/4-1) {\dot r_0\ov r_0} y^\a y^\a\ .
\ee
This transformation leaves the $SO(p-1)$ rotational subgroup
of the Lorentz group along
the brane invariant, as we will also see in the final expressions below.
Using the relation
\be
(\a+1)(d-2)=2\ ,\qq {\rm for}\quad M2, M5, D3, NS5\ ,
\ee
the metric takes the form
\be
 ds^2 = H_0^\a \Big[dy^\a dy^\a+
2 du dv + F(y,x,u) du^2 \Big] +  H_0^{1+\a} dx^i dx^i\ ,
\label{jsnnew}
\ee
with
\be
H_0=\sum_a {1\ov |{\bf x}-{\bf x}_a|^{d -2}}  \
\ee
and the profile function $F$ encoding the entire $u$-dependence is given by
\ba
F(y,x,u)& = &
\left({\ddot r_0\ov r_0} +\ha(d-6) {\dot r_0^2\ov r_0^2}\right)
\Big[\ha (d-4) y^\a y^\a + {2\ov d-4}
\sum_a{1\ov |{\bf x}-{\bf x}_a|^{d-4}}\Big]
\nonumber\\
&&  - {\dot r_0^2\ov r_0^2}
\sum_a {{\bf x}_a^2\ov |{\bf x}-{\bf x}_a|^{d-2}}\ .
\label{he2}
\ea
For $d=4$ the expression above is not valid and, as it turns out,
we should perform the replacement
\be
{2\ov d-4}
\sum_a {1\ov |{\bf x}-{\bf x}_a|^{d-4}}\to -2 \sum_a \ln |{\bf x}-{\bf x}_a|
\ .
\ee
As a check of the expression in \eqn{he2} recall that the equations of motion
are satisfied provided that $D^2F=0$, where the
Laplacian is defined with respect to the $D$-dimensional metric. This
translates to the condition
\be
H_0^{-1-\a}\del_\perp^2 F + H_0^{-\a}\del_\parallel^2 F=0 \ ,
\label{hr4}
\ee
where the subscripts in the two Laplacians indicate that they are taken
with respect to the $d$-dimensional and $(p-1)$-dimensional spaces perpendicular
and along the brane, respectively. Since this equation should be valid
for all $u$'s, the two terms in \eqn{he2} having
different $u$-dependences should satisfy it
separately. This is obvious for the term in the second line in \eqn{he2}.
The first line also satisfies \eqn{hr4} due to the relation
\be
\del_\perp^2 \sum_a{1\ov |{\bf x}-{\bf x}_a|^{d-4}} = -2(d-4)H_0
=-2 (d-4)\sum_{a}{1\ov |{\bf x}-{\bf x}_a|^{d-2}}\ .
\label{jfjw}
\ee
In the above manipulations we have used that
\be
\ha(d+\a D)=1+\a\ , \quad
(p-1)(d-4)=4\ ,\qq {\rm for}\quad M2, M5, D3\ .
\ee
In the case of NS5-branes when $d=4$ the necessary identity replacing
\eqn{jfjw} is
\be
\del_\perp^2 \sum_a \ln |{\bf x}-{\bf x}_a| =
2\sum_{a}{1\ov |{\bf x}-{\bf x}_a|^2}\ .
\ee

At this point we recall and make precise contact with
the solution generating technique of \cite{GarfVa}.
Accordingly, the construction of a new solution from some seed solution
with given matter content, requires the existence
of a null
Killing vector $\xi^\m$ satisfying the additional condition
\be
D_{[\m} \xi_{\n]}=\xi_{[\m} D_{\n]}S\ ,
\ee
for some scalar function $S$. If the right hand side vanishes then $\xi^\m$
is a covariantly constant null Killing vector.
A new solution with the {\it same}
matter fields can be constructed with metric
\be
G_{\m\n}+ e^S F \xi_\m \xi_\n\ ,
\label{jejs}
\ee
where we emphasize that the construction uses the Einstein frame metric.
The function $F$ satisfies
\be
\xi^\m D_\m F=0\ , \qq D^2 F = 0\ ,
\ee
that is it has vanishing Lie-derivative along the Killing vector and is
required to be harmonic. Other aspects and details
of this method have been analyzed in \cite{Kaloper}.
In our case the null vector is simply $\xi^\m\del_\m=\del/\del v$
and it turns out that the function $e^{-S}=H_0^\a$. Then
the deformed metric \eqn{jejs} gives precisely the line element in
\eqn{jsnnew}.
Note that even for the case of NS5-branes where there is a nontrivial dilaton
it turns that performing the computation in the Einstein frame and then
translating back into the string frame gives the correct result in
\eqn{jsnnew} with $\a=0$.
The method of \cite{GarfVa} was recently used in
\cite{Hubeny} to construct new solutions from
the single center branes of M- and string theory.
These solutions differ from ours (even in the single center cases)
due to the fact that they have as an essential part angular
dependence related to a spherical harmonic.

\subsection{The shock wave limit}

If the function $r_0(u)$ changes suddenly, say at $u=0$,
we expect to obtain
a shock wave geometry\footnote{The prototype example of
a gravitational shock wave geometry is that of a massless particle
moving in the four-dimensional Minkowski spacetime \cite{AiSexl}.}
since
such a change should affect the
geometry only in the null hypersurface at $u=0$.
However, it is not immediately clear how the
$\d$-function in the metric arises. If we simply demand that $\dot r_0\sim
\d(u)$ then the square of the $\d$-function appears and
we fail to make sense of it.
However, the gauge theory side of the correspondence gives an idea on how
to proceed. Recall again that the
centers of the harmonic function
correspond to vev's of scalar fields in the gauge theory.
From a physical view point a sudden vev change should be
associated with a finite energy pulse.
Such an energy is proportional to the integral of the
$T_{uu}$ component of the energy momentum
tensor along the $u$ direction, that is
\be
\int du \dot {\bf x}_a\cdot \dot{\bf x}_a \sim
\int du \dot r_0^2\ .
\ee
Therefore we should demand that
$\dot r_0^2$ behaves like a $\d$-function so that the integral
and consequently the energy of the pulse is finite.
Then, $\ddot r_0/r_0$
has a term behaving as $\d^\prime(u)$ which as a distribution can be
integrated by parts. With this procedure
we effectively replace $\ddot r_0/r_0$ by $\dot r_0^2/r_0^2$.
Taking all these into account in \eqn{he2}, we end up with the shock wave
profile
\be
F_{\rm shock}(y,x,u) =  V_{\rm shock}(y,x) \d(u)\ ,
\label{jproc}
\ee
with the non-trivial transverse space function
\be
V_{\rm shock}(y,x)
 =  f\ \Big({1\ov 4} (d-4)^2 y^\a y^\a +
\sum_a{1\ov |{\bf x}-{\bf x}_a|^{d-4}} -
\sum_a{{\bf x}_a^2 \ov |{\bf x}-{\bf x}_a|^{d-2}}\Big)\  ,
\ee
where the overall positive constant $f$ arises from $\dot r_0^2/r_0^2 =f\d(u)$.
In the case of $d=4$, corresponding to NS5-branes, only the last term survives
(the second term that becomes a constant can be absorbed by a harmless shift
of the variable $v$).

\no
A way to construct a shock wave solution in a
background geometry is by {\it cutting and pasting} a spacetime
along the $u=0$ hypersurface, that is by omitting the $\d$-function term,
replacing $v\to \hat v$
and
$dv\to d\hat v-\ha \Th(u) (\del_iV_{\rm shock} dx^i+\del_\a V_{\rm shock}
dy^\a)$, where $\Th(u)$ is the step function.
Changing variables as $ \hat v=v+\ha \Th(u) V_{\rm shock}$
we obtain back the standard form we 've been using.
This method was employed in order
to construct shock waves on purely gravitational
backgrounds in \cite{dray} and has
been generalized in the presence of matter fields, of a non-vanishing
cosmological constant as well as for shock waves
in string theory \cite{sfeshock1,sfeshock2}.\footnote{An alternative
method to obtain a shock wave is to start with a geometry in which the
deviation from Lorentz invariance is represented by a term containing a
small mass parameter.
By performing a infinite boost transformation and simultaneously
taking the mass parameter to zero in a correlated manner a shock profile
of the type \eqn{jproc} arises.
This method is very interesting from the physical point of view,
but at the same
time the cases where it can be applied are limited by the very
requirement that the unperturbed solution should be Lorentz invariant.
For notable applications and more details on this method
see \cite{AiSexl,dray} and \cite{Lousto}-\cite{Ortaggio}.} Also
shock wave solutions have been recently constructed in relation to
brane-world scenarios and brane-induced gravity \cite{Emparan,Kaloschock}.
Next, observe that the function $F_{\rm shock}$
depends (except for $d=4$)
not only on the transverse space coordinates $x^i$, but also on the
brane variables $y^\a$, via the rotational invariant combination $y^\a y^\a$.
This feature
has the consequence, as we will see, that momentum along the spatial
brane directions transverse to the $(u,v)$-lightcone will not be conserved
in scattering processes.

\no
An explicit realization for the function $r_0(u)$
with the desired properties is the following
\be
r_0(u)=a +{\e\ov 2}\tanh \left(lu\ov \e^2\right)\ ,
\label{ewjh}
\ee
where $a$, $l$ and $\e$ three parameter scales.
We see that for finite $u$ we have that $r_0(\pm\infty)=a\pm {\e\ov 2}$ which
in the limit of vanishing $\e$ implies that there is no change in $r_0$.
However, we easily check that this type of behaviour corresponds to
\be
\lim_{\e\to 0}{\dot r_0^2\ov r_0^2}= f  \d(u)\ ,\qq
f= { |l|\ov 3 a^2}\ .
\ee
With the representation \eqn{ewjh} one easily verifies that,
unlike its square, $\dot r_0/r_0$ is zero as a distribution.

Notice that, in the special case with
\be
r_0 \sim (\cosh u)^{d-4\ov 2}\ ,
\ee
giving rise to an exponential grow of $r_0$ in the far past and remote
future, all $u$-dependence in the metric disappears. Explicitly we have that,
up to a numerical constant
\be
F_{\rm exp}(y,x)=
{(d-4)^2\ov 4} y^\a y^\a + \sum_a{1\ov |{\bf x}-{\bf x}_a|^{d-4}}
 - \sum_a {{\bf x}_a^2\ov |{\bf x}-{\bf x}_a|^{d-2}}\ .
\ee

Finally we mention that, as it was shown for the case of shock waves
in \cite{sfeshock2}, the modifications of backgrounds we are considering,
can be given a string theoretical interpretation as marginal perturbations
by a massless
vertex operator along the lines discussed in the work of \cite{Callan}.

\section{Examples of varying brane distributions}
\setcounter{equation}{0}
\renewcommand{\theequation}{\thesection.\arabic{equation}}

Although we have kept the discussion so far as general as possible,
it is easier in practice to present some explicit examples in the
limit of continuous brane distributions since these have the
advantage to be describable by a finite number of moduli parameters.
In this section we consider the gravitational solutions
that arise from the shape invariant motion of a uniform continuous distribution
of D3-branes on a disc and on a three-dimensional spherical shell, both
with a $u$-dependent radii.
For the static case these solutions
were first constructed as the extremal limits of rotating D3-brane solutions
in \cite{trivedi,sfet1}. They were also used in several investigations
in the literature within the AdS/CFT correspondence starting with the works
of \cite{warn,brand} and belong to the rich class of
examples representing continuous distributions of
M- and string theory branes on higher dimensional ellipsoids
\cite{Basfe2}. It should be possible to construct the solutions
with moving moduli parameters in the more general cases as well.

\subsection{D3-branes on a disc of varying radius}

Consider the following parametrization of the transverse to the
D3-branes space $\IR^6$
\ba
\pmatrix{x_1\cr x_2} & = &\ r
 \cos\th \sin\psi \pmatrix{\cos\phi_2\cr \sin\phi_2}\ ,
\nonumber\\
\pmatrix{x_3\cr x_4} & = & \ r
\cos\th \cos\psi \pmatrix{\cos\phi_3\cr \sin\phi_3}\ ,
\\
\pmatrix{x_5\cr x_6} & = & \ \sqrt{r^2+1}
\sin\th \pmatrix{\cos\phi_1\cr \sin\phi_1}\ .
\nonumber
\ea
The ranges of the variables are
\be
0\le\th\ ,\psi<{\pi\ov 2} \ ,\qq
 0\le \phi_{1,2,3}< 2\pi\ ,\qq r\ge 0\ .
\label{je6}
\ee
In this parametrization the uniform D3-brane distribution occurs at the
$x_5\!-\! x_6$ plane, or for $\th=\pi/2$ and $r=0$, in a disc of unit radius.
This brane distribution breaks the $SO(6)$ $\cR$-symmetry to its $SO(4)\times
SO(2)$ subgroup, where the last factor is actually an approximation to a
discrete $Z_N$ group due to the continuum approximation.
The metric turns out to be
\ba
ds^2 & = & H^{-1/2} (2 du dv + dy_2^2+dy_3^2 + F du^2) + H^{1/2} {r^2+\cos^2\th
\ov r^2+1} dr^2
\nonumber\\
&& +\ H^{1/2}\left((r^2+\cos^2\th) d\th^2 + (r^2+1)\sin^2\th
d\phi_1^2 + r^2\cos^2\th d\Om^2_3\right)\ ,
\label{wn3}
\ea
where the harmonic function is (all integrals needed in this paper are
calculated with the aid of \cite{typologio})
\ba
H & = &\int_0^1 2 l dl \int_0^{2\pi}{d\phi\ov 2\pi}
 {1\ov (r_6^2+l^2-2 r_2 l \cos\phi)^2}=
\int_0^1 dl{2 l(r_6^2+l^2)\ov \left[(r_6^2+l^2)^2-4 r_2^2 l^2\right]^{3/2}}
\nonumber\\
&= &
{1\ov 2(r_6^2-r_2^2)}\left[1-{r_6^2-1\ov \sqrt{(r_6^2+1)^2-4 r_2^2}}\right]
= {1\ov r^2(r^2+\cos^2\th)}\
\ea
and the line element for the $S^3$ is explicitly given by
\be
d\Om^2_3= d\psi^2 + \sin^2\psi d\phi^2_2 + \cos^2\psi d\phi^2_3 \ .
\label{ejhw3}
\ee
We have also used the notation
\be
r_2^2=x_5^2+x_6^2= (r^2+1)\sin^2\th\ ,
\qq
r_6^2=x_1^2+\cdots + x_6^2= r^2+\sin^2\th\ .
\ee
The other sum that is needed can also be computed in the continuous
approximation
\ba
&& \sum_a {{\bf x}_a^2\ov |{\bf x}-{\bf x}_a|^4}  =   \int_0^1 2 l dl
\int_0^{2\pi} {d\phi\ov 2 \pi}
{l^2\ov (r^2_6+l^2-2 r_2 l \cos(\phi-\psi))^2}
\nonumber\\
&& =   2\int_0^{1} dl {l^3 (r_6^2+l^2)\ov [(r_6^2+l^2)^2-4 r_2^2 l^2]^{3/2}}
 =  {1\ov 2(r_6^2-r_2^2)}\left[r_6^2-{r_6^4+3 r_6^2 -4 r_2^2
\ov \sqrt{(r_6^2+1)^2-4 r_2^2}}\right]
\nonumber\\
&&
\ +\ln\left[
{r_6^2+1-2r_2^2+\sqrt{(r_6^2+1)^2-4 r_2^2}\ov 2(r_6^2-r_2^2)}\right]
\\
&&=
\ln\left(1+{1\ov r^2}\right)-{r^2-\sin^2\th\ov
r^2(r^2+\cos^2\th)}\ .
\nonumber
\ea
This can be checked to be a harmonic function in $\IR^6$. Finally,
we also have
the sum (computed also in the continuous approximation)
\ba
\sum_a {1\ov |{\bf x}-{\bf x}_a|^2}  & = &   \int_0^1 2 l dl
\int_0^{2\pi} {d\phi\ov 2 \pi} {l^2\ov r^2_6+l^2-2 r_2 l \cos(\phi-\psi)}
\nonumber\\
&= &    2\int_0^{1} dl{2 l\ov [(r_6^2+l^2)^2-4 r_2^2 l^2]^{1/2}}
\nonumber\\
&= & \ln\left[
{r_6^2+1-2r_2^2+\sqrt{(r_6^2+1)^2-4 r_2^2}\ov 2(r_6^2-r_2^2)}\right]
\\
&= &
\ln\left(1+{1\ov r^2}\right)\ .
\nonumber
\ea
Therefore the function $F$ in \eqn{wn3} is
\be
F={\ddot r_0\ov r_0}\left[y_1^2+y_2^2 +\ln\left(1+{1\ov r^2}\right)\right]
-{\dot r_0^2\ov r_0^2}\left[\ln\left(1+{1\ov r^2}\right)-{r^2-\sin^2\th
\ov r^2(r^2+\cos^2\th)}\right]\ .
\label{wkj1}
\ee
and for the case of a shock wave this is replaced by
\be
F_{\rm shock}= f \left[y_1^2+y_2^2 + {r^2-\sin^2\th\ov
r^2(r^2+\cos^2\th)}\right]\d(u)\ .
\label{wkj2}
\ee

\subsection{D3-branes on a sphere of varying radius}

Consider the following parametrization of the transverse to the
D3-branes space $\IR^6$
\ba
\pmatrix{x_1\cr x_2} & = &\ r
 \cos\th \sin\psi \pmatrix{\cos\phi_2\cr \sin\phi_2}\ ,
\nonumber\\
\pmatrix{x_3\cr x_4} & = & \ r
\cos\th \cos\psi \pmatrix{\cos\phi_3\cr \sin\phi_3}\ ,
\\
\pmatrix{x_5\cr x_6} & = & \ \sqrt{r^2-1}
\sin\th \pmatrix{\cos\phi_1\cr \sin\phi_1}\ ,
\nonumber
\ea
with the same ranges for the variables as in \eqn{je6}, except that now
$r\ge 1$.
In this parametrization the D3-brane distribution occurs at
$\th=0$ and $r=1$, in a spherical shell of of unit radius.
This brane distribution breaks the $SO(6)$ $\cR$-symmetry to its $SO(2)\times
SO(4)$ subgroup, where again the last factor is actually an approximation to a
discrete $Z_N$ group.
The metric turns out to be
\ba
ds^2 & = & H^{-1/2} (dy_1^2+dy_2^2+2 dudv + F du^2) + H^{1/2} {r^2-\cos^2\th
\ov r^2-1} dr^2
\nonumber\\
&& +\ H^{1/2}\left((r^2-\cos^2\th) d\th^2 + (r^2-1)\sin^2\th
d\phi_1^2 + r^2\cos^2\th d\Om^2_3\right)\ ,
\label{wn32}
\ea
where the harmonic function is
\ba
H  = {1\ov r^2(r^2-\cos^2\th)}
\ea
and the line element for the $S^3$ is explicitly given
by \eqn{ejhw3}.
Also in the continuous approximation
\be
\sum_a {{\bf x}_a^2\ov |{\bf x}-{\bf x}_a|^4}  =
-\ln\left(1-{1\ov r^2}\right)-{r^2+\sin^2\th\ov
r^2(r^2-\cos^2\th)}\ .
\nonumber
\ee
This can be checked to be a harmonic function in $\IR^6$. Finally, we also have
the sum (computed also in the continuous approximation)
\be
\sum_a {1\ov |{\bf x}-{\bf x}_a|^2} =
-\ln\left(1-{1\ov r^2}\right)\ .
\nonumber
\ee
Therefore the function $F$ in \eqn{wn3} is
\be
F={\ddot r_0\ov r_0}\left[y_1^2+y_2^2 -\ln\left(1-{1\ov r^2}\right)\right]
+{\dot r_0^2\ov r_0^2}\left[\ln\left(1-{1\ov r^2}\right)+{r^2+\sin^2\th
\ov r^2(r^2-\cos^2\th)}\right]\ .
\label{w1j1}
\ee
and for the case of a shock wave this is replaced by
\be
F_{\rm shock}= f \left[y_1^2+y_2^2 + {r^2+\sin^2\th\ov
r^2(r^2-\cos^2\th)}\right]\d(u)\ .
\label{w2j2}
\ee

\subsection{Shock waves on $AdS_p\times S^q$}

The most elementary examples one may construct are those for which we
place all brane centers at a single point, i.e. ${\bf x}_a=0$. Then
from \eqn{he2} we may compute the profile function for all cases of interest
and in particular for M2, M5 and D3 branes in the near horizon
for which the background geometry is given by the products
$AdS_4 \times S^7$, $AdS_7 \times S^4$ and $AdS_5 \times S^5$, respectively
\cite{gito}. We find the result
\ba
&& AdS_4\times S^7:\phantom{xxxx}
F=\ha \left({\ddot r_0\ov r_0} +{\dot r_0^2\ov r_0^2}\right)
\left[4 y_2^2 + {1\ov r^4}\right]\ ,
\nonumber\\
&& AdS_7\times S^4:\phantom{xxxx}
F= \left(2{\ddot r_0\ov r_0} -{\dot r_0^2\ov r_0^2}\right)
\left[{1\ov 4}(y_2^2+y_3^2 +y_4^2+y_5^2) + {1\ov r}\right]\ ,
\label{kj8}\\
&& AdS_5\times S^5:\phantom{xxxx}
F={\ddot r_0\ov r_0} \left[y_2^2+y_3^2 + {1\ov r^2}\right]\ .
\nonumber
\ea
However, for any smooth function $r_0(u)$ the fact that the profile function
$F$ is non-vanishing is an artifact of the coordinate system we are using.
Indeed, since all centers are at a single
point at the origin, there is no notion of a
shape invariant motion and we might as well use a coordinate system in which
$F=0$ without affecting the background geometry which remains simply
of the direct product type $AdS_p\times S^q$. However, let's consider the
shock wave limit. Then, the corresponding profile functions
$F_{\rm shock}$ are given by \eqn{kj8} by just replacing all
$u$-dependent prefactors by $f \d(u)$.
In that case the coordinate transformation becomes
extremely unwieldy both for mathematical computations and for preserving
the physical intuition since it involves the use of the step function and its
second power. This was already
noted for pure gravity shock waves in \cite{dray}.
We also remark that the shock waves on
$AdS_5\times S^5$ can be easily obtained from the expressions corresponding to
D3-branes on a disc, namely from \eqn{wn3}, \eqn{wkj1} and \eqn{wkj2}
by rescaling $r\to \l r$, $y_{1,2}\to y_{1,2}/\l$, $v\to v/\l^2$ and
then let $\l\to \infty$.
This procedure effectively sets the radius of the
disc to zero, thus restoring conformality.
A similar limiting procedure can also be performed
from the expressions corresponding to
D3-branes on a sphere, namely from \eqn{wn32}, \eqn{w1j1} and \eqn{w2j2},
leading to the same result.

\subsection{Field propagation and transitions}

It is possible to compute the amplitudes for transitions between different
eigenstates due to the presence of the shock wave, in field theory
by following a procedure similar to that for
the amplitude for scattering by a shock wave in four-dimensional Minkowski
spacetime computed in \cite{thooft}
(for further details and developments in relation also
to string theory see \cite{AmaCiaVene}).
Essentially one takes advantage of the
fact that the spacetime on either side of the shock wave is the same except
for the fact that the coordinate $v$ appears, as explained, shifted by
the function $V_{\rm shock}(y,x)$ defined in \eqn{jproc}.
For simplicity consider a scalar field $\Psi$.
For $u\to 0^-$ the solution has the form
\be
u\to 0^-:\qq \Psi_{k_v,{\bf n}}(v,x,y)=\Psi_{\bf n}(x,y) e^{ik_v v}\ ,
\ee
where $\Psi_{\bf n}$ denotes the complete set of
eigenfunctions solving the Laplace
equation in the transverse to the lightcone metric and
${\bf n}$ denotes collectively the corresponding set of quantum numbers.
For $u\to 0^+$ due the above shift of the coordinate $v$,
the solution should be
a linear combination of wavefunctions of the form
\be
u\to 0^+:\qq
\Psi_{k_v,{\bf n}}(v,x,y)=\Psi_{\bf n}(x,y) e^{ik_v (v+\ha V_{\rm shock})}\ .
\ee
Hence, the scattering amplitude is
\be
A_{k_v,{\bf n};k'_v,{\bf n'}}= \d(k_v-k'_v) A_{{\bf n};,{\bf n'}} \ ,
\label{hjamp}
\ee
where the $\d$-function expresses energy conservation and
the non-trivial part of the amplitude is
\be
A_{{\bf n};,{\bf n'}} = \int [dx dy] \Psi^*_{\bf n'}(x,y) \Psi_{\bf n}(x,y)
e^{{i\ov 2}k_v V_{\rm shock}}\ ,
\label{hhso2}
\ee
where the measure factor $[dx dy]$ includes the dilaton factor $e^{-2\Phi}$
if there is one.

\no
As an application,
consider a massless scalar propagating on $AdS_5\times S^5$.
Since from \eqn{kj8} the profile function of the shock wave
does not depend on the Euler angles parametrizing $S^5$ the quantum numbers
associated with $S^5$ will be conserved in the scattering process.
Hence, we focus on the $S$-wave solution which, when properly normalized,
reads
\be
\Psi(u,v,{\bf y},r) =
{1\ov (2\pi)^2} e^{i(k_u u + k_v v+ {\bf k}\cdot {\bf y})}
\Phi_M (z)\ ,\qq \Phi_M(z)= M^{1/2} z^2 J_2(M z)\ ,
\label{bell2}
\ee
with $z=1/r$, ${\bf y}=(y_1,y_2)$, ${\bf k}=(k_1,k_2)$ and
$M^2=-k\cdot k=-2 k_u k_v-{\bf k}^2$.
Also $J_2$ is the Bessel function of index 2,
which is regular at $z=0$.
The arbitrary overall constant is chosen so that the Dirac-type
normalization condition is satisfied
\be
\int_0^\infty {dz\ov z^3} \Phi_M \Phi_{M'} = \d(M-M')\ .
\label{noormm}
\ee
Then the amplitude from a state with $(k_u,k_v,{\bf k})$ to
$(k'_u,k'_v,{\bf k'})$ is given by \eqn{hjamp} with
\ba
A_{k_u,{\bf k};k'_u,{\bf k'}}
 & = & {\sqrt{M M'}\ov 2\pi}\int_{-\infty}^{+\infty} d^2{\bf y}e^{i({\bf k}
-{\bf k'})\cdot {\bf y}} e^{{i\ov 2}k_v f {\bf y}^2}
\int^\infty_0 dz z J_2 (Mz)J_2 (M'z) e^{{i\ov 2}k_v f z^2}\ .
\nonumber\\
&=& -{\sqrt{MM'}\ov  k_v^2 f^2} e^{i{k\cdot k'\ov k_v f}}
J_2\left({MM'\ov k_v f}\right)\ .
\ea
Note the absence of momentum conservation along the two brane coordinates
transverse to the lightcone. This is due to the dependence
of the shock wave profile on these coordinates.
The fact that one can compute the scattering amplitude off a shock wave on
$AdS_5\times S^5$ explicitly, suggests the interesting possibility to further
explore the r\^ole of shock waves within the AdS/CFT correspondence.
Work on that relation has appeared in the literature \cite{itzaki}.
We believe that much more remains to be understood.

\section{NS5-branes on a circle}
\setcounter{equation}{0}
\renewcommand{\theequation}{\thesection.\arabic{equation}}

In this section we consider NS5-branes with
centers on a $N$-polygon situated
on the $x_3$-$x_4$ plane in the $R^4$ space transverse to the
branes. Following \cite{sfet1} we have
\be
{\bf x}_a = (0,0,\cos\phi_a ,\sin\phi_a)\ ,\qq \phi_a= 2\pi{a\ov N}\ ,
\quad a=0,1,\dots , N-1\ .
\ee
This distribution of branes preserves an $SO(2)\times Z_N$ subgroup
of the $SO(4)$ original $\cR$-symmetry group when all branes are located at a single
point of the transverse space.
We will be interested in motions of the centers of the branes preserving
this symmetry. Hence, we allow the parameter $r_0$ to depend on $u$ which is
a radial motion that preserves the angular distance between the branes.
In the continuum limit the branes are distributed on a ring of varying
radius $r_0(u)$ situated at the (34)-plane and the subgroup of the
$\cR$-symmetry preserved by
our configuration becomes continuous, i.e. $SO(2)\times SO(2)$.
After changing variables as
\be
\pmatrix{x_1\cr x_2}= \sinh\r \cos\th \pmatrix{\cos\tau\cr \sin\tau}\ ,\qq
\pmatrix{x_3\cr x_4}= \cosh\r \sin\th \pmatrix{\cos\psi\cr \sin\psi}\ ,
\ee
with ranges
\be
0\le \r<\infty \ , \qq 0\le \th < {\pi\ov 2}\ ,\qq 0\le \psi,\tau< 2\pi\ ,
\ee
we find that\footnote{Even in the discrete case it is possible to explicitly
compute $H$ \cite{sfet1}.}
\be
H =  {1\ov \sqrt{(x_1^2+x_2^2+x_3^2+x_4^2+r_0^2)^2-4 r_0^2 (x_3^2+x_4^2)}}
=  {1\ov \sinh^2 \r + \cos^2\th}\ .
\ee
In this parametrization the ring is at $\rho =0$ and $\th={\pi/2}$.
After a straightforward computation we find that
the 6-dim non-trivial part of the background is
\ba
ds^2 & = &  2 dudv
+ d\r^2 + d\th^2 +{\tan^2\th d\psi^2 + \tanh^2 \r d\tau^2\ov 1+
\tan^2\th \tanh^2\r} + F du^2 \ ,
\nonumber\\
B_{\tau\psi}& = & {1\ov 1+\tan^2\th \tanh^2\r} \ ,
\label{gg1}\\
e^{-2 \Phi}& = & r_0^2(u) (\sinh^2\r + \cos^2\th)\ ,
\nonumber
\ea
where
\be
F  =   - \left(\dot r_0\ov r_0\right)^2 {1\ov \sinh^2\r+\cos^2\th}
-2 {d\ov du}\left(\dot r_0\ov r_0\right)\ln \cosh \r\ .
\label{Pp2}
\ee
In the case of the shock wave this should be replaced by
\be
F_{\rm shock}  =   - f {1\ov \sinh^2\r+\cos^2\th}\ \d(u)\ .
\label{Pp2sh}
\ee
and the $r_0^2(u)$ factor in the dilaton is set to a constant.
In the static case, when the function $F=0$, it was shown in \cite{sfet1}
that a T-duality transformation with respect to $\tau$, relates
the background corresponding to NS5-branes on a ring to the
background for the $SL(2,\IR)/U(1)\times
SU(2)/U(1)$ exact CFT (actually an orbifold of it, see \cite{Giveon} and
for further details \cite{Israel}). In the non-static case,
performing a T-duality  transformation with respect to $\tau$ we find
\ba
ds^2 & = & 2 dudv + d\r^2 + \coth^2 \r d\om^2
+ d\th^2 +\tan^2\th d\tau^2 + F du^2
\nonumber\\
e^{-2 \Phi}& = & r_0^2(u) \cos^2\th \sinh^2\r\ ,
\label{gg3}
\ea
where $\om= \tau+\psi$ and zero antisymmetric tensor.
For the case of a shock wave we simply replace
$F$ by $F_{\rm shock}$ and set $r_0(u)$ to a constant.

\subsection{Generalities on transitions}

To see the effect of the time-changing moduli we consider
a scalar field propagating in the geometry \eqn{gg1}.
Since we would like our formalism to be applicable
to more general cases, let's consider a general transverse metric.
In addition, we will develop the formalism for a general profile function
which will be suitable for various approximations schemes.
When we specialize to the case of a shock wave
it is possible to obtain the exact amplitude we have already seen.
In particular, we let our string frame metric be of the form
\be
ds^2 =  dy^2_1+\cdots + dy_4^2 + 2 du dv + F(u,x)du^2
+ g_{ij}(x) dx^i dx^j \ ,\qq i=1,2,3,4\ .
\ee
Then the standard massless wave equation reads
\be
{1\ov e^{-2 \Phi} \sqrt{-G} } \del_\m e^{-2 \Phi} \sqrt{-G}  G^{\m\n}
\del_\n \Psi =0\ ,
\label{hjd}
\ee
where we note that if we write it using the Einstein frame metric the
dilaton factor does not appear. Making the ansatz
\be
\Psi(u,v,y,x)={1\ov (2\pi)^{5/2}} e^{i \vec k\cdot \vec y}
e^{i k_v v} \Psi(u,x)\ ,
\ee
we derive an equation for the amplitude $\Psi(u,x)$ written in the
suggestive form
\be
(H^{(0)}+H^{(1)})\Psi = -2 ik_v \dot \Psi\ ,
\label{je9}
\ee
where
\be
H^{(0)}= {1\ov e^{-2 \Phi} \sqrt{g} }
\del_i e^{-2 \Phi} \sqrt{g} g^{ij}\del_j -\vec k^2\ ,
\ee
denotes the unperturbed Hamiltonian and
\be
H^{(1)}=k_v^2 F(u,x) -2 i k_v \dot \Phi\ ,
\label{hh11}
\ee
is the, not necessarily small, perturbation.
Let $\{\Psi_n^{(0)}\}$  be a complete set of
states that solve the unperturbed problem, with $n$ denoting collectively
all quantum numbers. They have the form
\be
\Psi^{(0)}_n (u,x )= {1\ov (2\pi)^{1/2}}
\Psi^{(0)}_n (x ) e^{-i {E_n+\vec k^2\ov 2 k_v}u}
\ee
and the amplitude solves the equation
\be
{1\ov e^{-2 \Phi} \sqrt{g} }
\del_i e^{-2 \Phi} \sqrt{g} g^{ij}\del_j \Psi_n^{(0)} + E_n\Psi_n^{(0)}= 0\ ,
\label{dj01}
\ee
defined completely in the four-dimensional transverse space.
Proceeding further we expand the solutions of \eqn{je9} as
\be
\Psi(u,x)=\sum_n a_n(u) \Psi^{(0)}_n(u,x)\ .
\ee
Substituting into \eqn{je9} we derive a system of coupled first order
equations given by
\be
\dot a_m(u)={i\ov 2 k_v} \sum_n
e^{i\om_{mn} u'} H^{(1)}_{mn} a_n(u)\ ,\qq \om_{mn}={E_m-E_n\ov 2 k_v}\ ,
\label{jhf}
\ee
where the matrix elements are\footnote{
If the system is initially at the $i$-th state and
$H^{(1)}$ can be treated as a perturbation, then the coefficients
$a_m$, with $m\neq i$ are small compared to $a_i\simeq 1$. Hence,
to first order in perturbation theory
\be
a^{(1)}_m(u)= {i\ov 2 k_v} \int_{-\infty}^u du'
e^{i\om_{mi} u'} H^{(1)}_{mi}(u')\ \ ,\qq m\neq i\ .
\label{kjl1}
\ee
If we are interesting in the final state we just let $u\to \infty$ in the
above integral.
}
\be
H^{(1)}_{mn}(u)
=\int d^4x e^{-2\Phi}\sqrt{g}\Psi_m^{(0)*}(x) H^{(1)}\Psi^{(0)}_n(x)\ .
\label{kkk2}
\ee
In the special case of shock waves we have using \eqn{jproc} that
\be
H^{(1)}=k_v^2  \d(u) V_{\rm shock}(x)\
\label{ejhw}
\ee
and the amplitude can be exactly computed in this more general approach
leading to the same results we have mentioned before.
Indeed, when $H^{(1)}$ is of the form \eqn{ejhw}, the system \eqn{jhf} becomes
\be
\dot a_m(u)={ik_v f\ov 2} \sum_n (V_{\rm shock})_{mn} a_n(u) \d(u)\ ,
\label{jhfsc}
\ee
where $(V_{\rm shock})_{mn}$ is given by an expression similar to \eqn{kkk2}.
This system can be readily solved giving
\be
a_m =\int d^4x e^{-2\Phi}\sqrt{g}\Psi_m^{(0)*}(x) \Psi^{(0)}_i(x)
e^{ {i\ov 2}k_v  V_{\rm shock}(x)} \ , \qq {\rm for}\quad u>0\ .
\ee
This is indeed of the form \eqn{hhso2} with the measure $[dxdy]$ being
$d^4 x e^{-2\Phi}$, since there is no longitudinal contribution to the shock
wave profile and the dilaton field has been properly taken into account.

\subsection{The spectrum}

In order to proceed we need to compute the eigenfunctions and spectrum
in \eqn{dj01}. For our static background \eqn{gg1} (with $F=0$)
this is possible to do by the method of separation of variables. Let
\be
\Psi^{(0)}(\rho,\th,\psi,\tau)={1\ov 2\pi}e^{im\psi}e^{i n \tau}
\Th (\th) R(\r)\ .
\ee
In this way, after standard manipulations,
we obtain two ordinary linear second order differential equations
\be
{1\ov \sin 2\th}{d\ov d\th}\left(\sin 2\th {d\Th\ov d\th}\right)
+ \left(E_1 -m^2 \cot^2\th -n^2 \tan^2\th \right)\Th=0
\label{diif1}
\ee
and
\be
{1\ov \sinh 2\r}{d\ov d\r}\left(\sinh 2\r {dR\ov d\r}\right)
+ \left(E_2 -m^2 \tanh^2\r -n^2\coth^2\r \right)R=0\ ,
\label{diif2}
\ee
where $E_1$ and $E_2$ arise as separation of
variables constants and obey $E_1+E_2=E$, where $E$ is the energy eigenvalue.

\no
The solution to \eqn{diif1} is given in terms
of Jacobi Polynomials as
\be
\Th_{l,n,n}(\th) =  A_{l,m,n}\sin^{|m|}{\th}\cos^{|n|}{\th}
P^{(|m|,|n|)}_{l-{|m|\ov 2}-{|n|\ov 2}}(\cos 2\th)\ ,\qq
 l-{|m|\ov 2}-{|n|\ov 2}=0,1,\dots \ ,
\label{doo3}
\ee
where the normalization constant is
\be
A^2_{l,m,n} = (2l+1) {\G(l+\ha |m|+\ha |n|+1)
\G(l-\ha |m|-\ha |n|+1)\ov \G(l+\ha |m|-\ha |n|+1)
\G(l-\ha |m|+\ha |n|+1)}\ ,
\label{noorm3}
\ee
so that the state is normalized to one w.r.t. the measure $d\th \sin2\th$.
The spectrum is quantized accordingly as
\be
(E_1)_{l,m,n}= 4 l(l+1)-m^2 - n^2 \ .
\label{eenn1}
\ee
We note here that $l$ is generally half an integer
and for the particularly interesting case with $m=n=0$ we have
\be
\Th_l(\th) =  \sqrt{2 l+1}\ P_{l}(\cos 2\th)\ ,\qq
 l=0,1,\dots \ ,
\label{doo3l}
\ee
where $P_l$ denote the Legendre polynomials.

\no
The differential equation \eqn{diif2} can be cast by an appropriate
transformation into a hypergeometric
differential equation.
The particular form of the solution depends very much on the values
of $M_2^2$. We find convenient to parametrize
\be
(E_2)_{j,m,n}= m^2+n^2 -4 j(j+1)\ ,
\label{eenn2}
\ee
so that the total energy is
\be
E_{l,j} = 4 l(l+1)-4 j(j+1)\ .
\label{eenn}
\ee
The solution for $R(\r)$ is given in terms of hypergeometric functions.
There are different ways of representing it.
If we let
\be
R= x^{|m|/2} (x-1)^{|n|/2} F(x) \ ,
\label{ccc1}
\ee
where $x=\cosh^2{\r}$, we find that the function $F(x)$ obeys
the hypergeometric equation with
\be
a={|m|\ov 2}+{|n|\ov 2}+j+1\ ,\qq b={|m|\ov 2}+{|n|\ov 2}-j\ ,\qq c=1+ |m|\ ,
\ee
in the standard notation.

\no
If we let $z=1/\cosh^2 \r$ and make the transformation
\be
R= z^{j+1}(1-z)^{|n|/2} F(z)\ ,
\label{ccc2}
\ee
we obtain for $F$ a hypergeometric differential equation with
\be
a={|m|+|n|\ov 2}+j+1\ ,\qq b={|n|-|m|\ov 2}+j+1\ ,\qq c=2(j+1)\ ,
\ee
in the standard notation.
If we make instead the transformation
\be
\psi_2= z^{-j} (1-z)^{|n|/2} F(z)\ ,
\label{ccc3}
\ee
we obtain for $F$ a hypergeometric differential equation with
\be
a={|m|+|n|\ov 2}-j\ ,\qq b={|n|-|m|\ov 2}-j\ ,\qq c=-2j\ .
\ee
The latter cases are related by the replacement $j\to -j-1$.

\no
The different solutions are characterized by the values of $j,m$
and $n$. We will be particularly interested in the case where
$j=i\s -\ha$. In that case we will obtain an unbound solution that behaves
as a plane wave asymptotically.
Using \eqn{ccc2} we find the solution (up to a normalization constant)
\be
R_{\s,m,n}(\r) = z^{1/2} (1-z)^{|n|/2}\left( z^{i\s} e^{i\varphi}
{}_2F_1(a,b,c,z) +c.c.\right)\ ,\qq z={1\ov \cosh^2{\r}}\ ,
\ee
where
\ba
&&a=\ha(|m|+|n|+1) + i\s\ ,\qq b=\ha(|n|-|m|+1) +i\s\ ,\qq c=1+2 i\s\ ,
\nonumber\\
&& e^{2 i\varphi}={\G(-2 i\s)\Gamma(a)\Gamma(b)\ov
\G(2 i\s)\Gamma(a^*)\Gamma(b^*)}\ .
\ea
The relative coefficient between the two terms has been fixed so that
the solution is regular at $z=1$ (corresponding to $\r=0$).
The energy eigenvalue is given as
\be
E_{l,\s}=4l(l+1)+4\s^2+1\ ,
\ee
corresponding to a continuous spectrum that has a mass gap $E_{\rm gap}=1$
and a discrete part superimposed on it.

\no
An equivalent way of representing the solution can be found using properties of
the hypergeometric functions
\ba
&& R_{\s,m,n}(\r)= \sqrt{{2\s\ov \pi}\sinh 2\pi\s}
{\left|\G(a)\G(b)\right|\ov |n|!}
 \cosh^{|m|}{\r}\sinh^{|n|}{\r}\
{}_2\!F_1(\a,\a^*,1+ |n|,-\sinh^2{\r})\ ,
\nonumber\\
&& \a={|m|+|n|+1\ov 2}+i\s\ ,
\ea
where $\s$ is real number.
We have the limiting behaviours
\be
R_{\s,m,n}(\r) \simeq \sqrt{{2\s\ov \pi}\sinh 2\pi\s}
{\left|\G(a)\G(b)\right|\ov |n|!} \r^{|n|} \ ,\qq {\rm as}\quad \r\to 0\ .
\ee
and
\be
R_{\s,m,n}(\r)\simeq 2   \left( 4^{i\s}e^{-i(2\s\r-\varphi)}
+ c.c.\right) e^{-\r} \ ,\qq {\rm as}\quad \r\to \infty\ .
\ee
Therefore, due to the behaviour at infinity, these are unbound states
well suited for scattering problems.

\no
A particularly interesting case in which the various integrals
appearing below in the evaluation of scattering amplitudes
will be
possible to explicitly compute, is when the quantum number $\s\gg 1$. Then
since the corresponding energy $E_2$ is very high the solutions simplifies
to that for a plane wave (for notational convenience we let $\s\to \s/2$)
\be
R_{\s}(\r)= \sqrt{2\ov \pi \sinh 2 \r}\ \cos \s \r\ ,\qq \s>0 \ ,\quad \r>0\ .
\label{mnre}
\ee
so that they form a complete orthonormal set in $\IR^+$ for $\r$ and $\s$,
with measure $\sinh 2\r$.

\subsection{Computing the transition amplitude}

Returning to our case and concentrating to scattering by the shock wave,
we have from \eqn{Pp2sh}, \eqn{hh11} and \eqn{ejhw} that
\be
V_{\rm shock}(\r,\th)= - {1\ov \sinh^2\r + \cos^2\th}\ .
\label{dij6}
\ee
It is clear that in computing the amplitude we necessarily have that the
quantum numbers $m$ and $n$ do not change. Below we are interested in
cases where $m=n=0$, $l={\rm integer}$
and $\s\gg 1$ since then our computations can be
performed most easily. We will denote the initial state in the remote
past by the quantum numbers $l'$ and $\s'$ and the corresponding unprimed
quantities $l$ and $\s$ will denote the quantum numbers in the remote
future.
Then the relevant wavefunctions for the computation of the amplitude
are given by \eqn{doo3l} and \eqn{mnre}. We have in general that the
exact amplitude is given by
\be
a_{l,\s;l'\s'}
  =   {2 c_{l}c_{l'}\ov \pi}
\int_0^{\infty}d\r \cos (\s \r) \cos (\s' \r) \int_0^{\p/2}
{d\th \ \sin 2\th }
e^{ {i\ov 2} k_v f V(\r,\th)}
P_l(\cos 2\th) P_{l'}(\cos 2\th)\ ,
\label{jk9}
\ee
where $c_l=(2l+1)^{1/2}$.

\subsubsection{Low frequency perturbative expansion}

For $k_v f\ll 1$ we may expand the phase factor and to first order in
perturbation theory we compute\footnote{To
perform the various integrations in \eqn{jk9} and \eqn{dj8} below
we have used \cite{typologio} and in particular eqs.
7.224(5), 8.825 and 3.983(1).}
\ba
a^{(1)}_{l,\s;l'\s'} & = &
-i{k_vf\ov \pi} c_lc_{l'} \int_0^\infty d\r \cos\s\r \cos\s'\r
\int_0^{\pi/2} {d\th\ \sin 2\th\ov \sinh^2\r + \cos^2\th}
P_l(\cos 2\th) P_{l'}(\cos 2\th)
\nonumber\\
& = & -i{2 k_v f\ov \pi}  (-1)^{l+l'} c_l c_{l'}
\int_0^\infty d\r \cos(\s \r) \cos(\s'\r) Q_l(\cosh 2 \r) P_{l'}(\cosh 2\r)\ ,
\ea
where in order to for the second line to be valid
we have assumed with no loss of generality that $ l'\le l$.
We may further simplify the expression for $a^{(1)}_{l,\s,l',\s'}$
when $l'=0$
corresponding to the $S$-wave as far as the angular part is concerned.
Then
\ba
a^{(1)}_{l,\s,0,\s'} & = &-i {2k_v f\ov \pi} (-1)^l c_l
\int_0^\infty d\r \cos (\s \r) \cos (\s'\r) Q_l(\cosh 2 \r)
\nonumber\\
& = & - i {k_v f\ov \pi} c_l
 \int_{-1}^1 dt P_n(t) \int_0^\infty d\r {\cos (\s\r) \cos(\s'\r)
\ov \cosh 2 \r+t}
\label{dj8}\\
& = & -i {k_v f\ov 4 } c_l
\left(\int_0^\pi d\phi
P_l(\cos\phi) {\sinh(\s_- \phi)\ov \sinh (\pi \s_-)}+(\s_-\to \s_+)\right)\ ,
\nonumber
\ea
where in the last line we have changed the integration variable as
$t=\cos\phi$ and used the definition $\s_\pm =\ha (\s\pm\s')$.
Recall that both $\s,\s'\gg 1$ and therefore $\s_-$ could be
finite, but  necessarily $\s_+\gg 1$.
Hence, although we have kept the corresponding term
this in fact should be neglected as we do from now on.
The amplitude drops as $1/\s_-$, for $\s_-\gg 1$
and remains finite when $\s_-\to 0$.
For low values of $l$ we have explicitly that
\ba
a^{(1)}_{0,\s,0,\s'} & = &  -i {k_v f \ov 4}
{1\ov \s_-}
\tanh\left({\pi\ov 2}\s_-\!\right)\ ,
\nonumber\\
a^{(1)}_{1,\s,0,\s'} & = &  i {\sqrt{3}k_v f \ov 4}
{\s_- \ov 1+\s_-^2} \coth\left({\pi\ov 2}\s_-\!\right)\ ,
\\
a^{(1)}_{2,\s,0,\s'} & = &  -i {\sqrt{5}k_v f \ov 4}
{1+\s_-^2 \ov \s_- (4+\s_-^2)}\tanh\left({\pi\ov 2}\s_-\!\right)
\ .
\nonumber
\ea

\subsubsection{High frequency asymptotic expansion}

In this case we have the opposite
limit with $k_v f\gg 1$ and the phase factor in \eqn{jk9} oscillates very
rapidly. We may evaluated the leading contribution in the saddle point
approximation. To do that note that the integral over the $\r$ variable,
after changing variables as $x=e^{-2 \r}$, becomes
\be
{1\ov 8} \int_0^1 d x (x^{-{i\ov 2} \s}+ x^{+{i\ov 2} \s})
(x^{-{i\ov 2} \s'}+ x^{+{i\ov 2} \s'}) e^{i k_v f\psi(x)}\ ,\qq
\psi(x)={-2\ov x+x^{-1} + 2\cos2\th}\ .
\ee
The phase factor $\psi(x)$ has a stationary point at $x=1$, where
$\psi^\prime(1)=0$. Using $\psi''(1)=1/(4 \cos^2\th)$ we obtain that
asymptotically the integral behaves as
\be
\sqrt{\pi\ov 2 |k_v f|}\ {1\ov \cos^2\th} \ e^{\pm {i \pi\ov 2}}
e^{-{i k_v f\ov 2 \cos^2\th}}\ ,
\ee
where the sign in the exponent is that same as the sign of $k_v f$.
Next we find the leading behaviour of the remaining integral over $\th$.
After changing integration variable as $x=\cos 2\th$ we have to consider the
integral
\ba
&& \int_{-1}^1 {dx\ov x+1} P_l(x)P_{l'}(x) e^{-i {k_v f\ov x+1}}
 = {1\ov i k_v f} \int_{-1}^1 dx (x+1) P_l(x)P_{l'}(x)
{d\ov dx}e^{-i {k_v f\ov x+1}}
\nonumber\\
&& =  {2\ov i k_v f}  e^{-{i\ov 2} k_v f} + {\cal O}\left(
1\ov k_v f\right)^2\ ,
\ea
where we simply performed an integration by parts
and have used that $P_l(1)=1$ for the Legendre polynomials.
Putting everything together we find that the amplitude behaves asymptotically
as
\be
a_{l,\s,l',\s'}\approx \sqrt{8\ov \pi} \sqrt{2 l+1}\sqrt{2 l'+1} {1\ov
|k_v f|^{3/2}} e^{-{i\ov 2} k_v f}\ .
\ee
The amplitude in this limit does not depend on the quantum
numbers $\s$ and $\s'$.

\section{Open strings ending on moving branes}
\setcounter{equation}{0}
\renewcommand{\theequation}{\thesection.\arabic{equation}}

In this section we will consider an open string with one end attached
on a fixed $Dp$-brane and the other on a moving $Dp$-brane.
Our aim is to find a possibly generic mechanism which might be responsible
for producing at the macroscopic level a shock wave on the gravitational
background of a large number of branes.
Our treatment is parallel to that
of an open string interacting with its ends with a wave of arbitrary profile
\cite{Bachas}. We also note that the philosophy and some of the mathematical
techniques we will use have appeared in the present context some time ago
in order to demonstrate the solvability of the first quantized string
in shock wave backgrounds of arbitrary profile
on Minkowski space-time \cite{Amati}.
We will consider a $d$-dimensional flat space-time
with Minkowski signature so that the directions along the branes satisfy
the usual NN boundary conditions.
For the coordinates normal to the branes we have to choose DD boundary
conditions. We will take the moving brane to have a time-dependent position
via the light-cone coordinate $u=X^0+X^1$.
With this choice it is possible
to solve the string equations of motion in the light-cone gauge in which
$u=\tau$.
In this case the other light-cone variable $v$ is determined
as usual in terms of the remaining $d-2$ transverse coordinates
denoted by $X^i$, using the Virasoro constraints.
We split the index $i=(a,I)$ where $a$ and $I$ refer to the directions along
and normal to the brane, respectively.
We will take the spatial world-sheet coordinate $\s\in [0,l]$.
Hence, we may easily take the length of the string to zero, i.e. $l\to 0$,
corresponding to the point particle limit.
Then, we have the following boundary conditions
\be
NN: \qq\qq
\del_\s X^{a}(\tau,0)=\del_\s X^{\m}(\tau,l)= 0\ ,\qq a=2,\dots , p\ .
\ee
and
\be
DD: \qq\qq
X^I(\tau,0)=0\ ,\quad X^I(\tau,l)= A^I + f^I(\tau)\ ,\qq I=p+1,\dots , d\ ,
\label{hdh}
\ee
where $A^I$ are constant vectors and
where $f^I(\tau)$ is a given set of function assuming to behave in the
far past and future as\footnote{Before
the light-cone gauge choice is made we may use
$f^I(u(\tau,l))$, instead of $f^I(\tau)$.}
\be
f^I(-\infty)=0\ ,\qq f^I(+\infty)=B^I \ .
\label{jkdf}
\ee
Therefore the ends of the string are stretched in the $I$th direction
a distance $A^I$ in the far past,
and a distance $A^I+B^I$ in the far future.
The two-dimensional action for the transverse coordinates is
\be
S={1\ov 2 l}\int^l_0 d\tau d\s \left(\del_\tau X^i \del_\tau X^i +
\del_\s X^i \del_\s X^i \right)\ .
\ee
The solution of the equations of motion
\be
\del_\tau^2 X^i - \del_\s^2 X^i = 0 \ ,
\label{ewrh}
\ee
for the longitudinal coordinates is simply given by
\be
X^a(\tau,\s)=  x_0^a + a_0^a \tau +i \sqrt{l\ov \pi}
 \sum_{n\neq 0} {a^a_n\ov n} \cos(n\pi \s/l)e^{-i n\pi \tau/l}
\ .
\label{kjs1}
\ee
Using the general equal time computation relations
\be
[X^i(\tau,\s),X^j(\tau,\s')]=[P^i(\tau,\s),P^j(\tau,\s')]=0\ ,
\qq [X^i(\tau,\s),P^j(\tau,\s')]=i\d^{ij}\d(\s-\s')\ ,
\ee
where in the momentum is defined as $P^i={\dot X^i\ov 2l}$,
we find the usual commutation
algebra for the longitudinal mode coefficients
\be
[a^a_n,a^b_m]=n \d_{n+m} \d^{ab}\ ,\qq [x_0^a,p_0^b]=i \d^{ab}\ ,
\ee
with $p_0^a=a_0^a/l$ the zero mode momentum.
For the transverse to the brane coordinates we write
\be
X^I(\tau,\s)=x_0^I(\tau,\s) + \bar X_0^I(\tau,\s)\ ,
\label{hjein}
\ee
where
$x_0^I$ is a classical piece that satisfies \eqn{ewrh}
and the DD boundary condition \eqn{hdh} with $A^I=0$.
In addition, we demand that it obeys the initial condition
\be
\lim_{\tau\to -\infty} x_0^I(\tau,\s)=0\ .
\label{f46}
\ee
The $\bar X^I$'s represent the fluctuations around $x_0^I$
and satisfy the usual DD boundary condition
$\bar X^I(\tau,0)=0$ and $\bar X^I(\tau,l)=A^I$.
The most general solution for the fluctuations is
\be
\bar X^I(\tau,\s)= {\s\ov l}A^I + \sqrt{l\ov \pi}
 \sum_{n\neq 0} {a^I_n\ov n} \sin(n\pi \s/l)e^{-i n\pi \tau/l}\ .
\label{fluui}
\ee
For the classical part we find that the solution is given by
\be
x^I_0(\tau,\s)
= \sum_{n=1,3}^\infty \left[f^I(\tau+\s-n l)-f^I(\tau-\s-n l)\right]\ ,
\label{j2fg1}
\ee
where the sum extends over the positive odd integers.
Clearly \eqn{j2fg1} satisfies the equation of motion \eqn{ewrh} and the
appropriate boundary condition \eqn{hdh} (with $A^I=0$). In addition, due
to the
behaviour \eqn{jkdf} it vanishes in the far past as demanded by \eqn{f46}.
In the far future manipulations with the terms in the infinite sum are
problematic since each term separately diverges. However, an appropriate
regularization procedure yields the expected
result $x^I_0(+\infty,\s)=B^I\s/l$. Therefore in the far future the solution is
given by \eqn{fluui} with $A^I$ replaced by $A^I+B^I$.
The classical expression \eqn{j2fg1}
contains no free moduli parameters to be quantized and the
equal time computation relations
give rise to the algebra for the mode coefficients of the fluctuations
\be
[a^I_n,a^J_m]=n \d_{n+m} \d^{IJ}\ .
\ee
This define an ``in'' vacuum and the question is whether
a unitary operator $U(\tau)$ exists,
that evolves the ``in'' solution to the
full solution \eqn{hjein}. Namely that
\be
X^I(\tau,\s)=x_0^I(\tau,\s) + \bar X^I(\tau,\s) = U\inv(\tau) \bar X^I(\tau,\s)
U(\tau)\ .
\label{hjd3}
\ee
Given the initial condition \eqn{f46} we demand that $U(-\infty)=1$.
Then, the $S$-matrix describing the unitary evolution to the final state
is by definition given by
\be
S=U(\infty)\ .
\ee
For the unitary operator $U(\tau)$ we make the ansatz
\be
U(\tau)=e^{iA(\tau)} \ ,\qq A(\tau)= \sum_n \l_n^I(\tau)a_n^I\ ,\quad
(\l^I_{-n})^*=\l^I_n\ ,
\ee
where the last condition ensures the hermiticity of $A$.
In order to compute the coefficients $\l_n^I$ we use \eqn{hjd3} and the fact
that the commutator $[A,\bar X^I]$ is a $c$-number. We immediately find that
\ba
x_0^I(\tau,\s) &=& -i [A(\tau),\bar X^I(\tau,\s)]
= -i  \sum_n \l_n^I(\tau) [a_n^I,\bar X^I(\tau,\s)]
\nonumber\\
& = &
 -i\sqrt{l\ov \pi} \sum_n \l_n^I(\tau)\sin{n\pi\s\ov l}
e^{i n\pi \tau/l}\ .
\ea
Using for the left hand side \eqn{j2fg1} we may express
\be
\l_n^I(\tau)=-\sqrt{\pi\ov l^3}
\int_0^{2 l}d\s e^{-i n \pi (\tau+\s)/l}
\sum_{m=1,3} f^I(\tau+\s-m l)\ .
\ee
Using the identity
\ba
\sum_{m=1,3} f^I(\tau+\s-m l) & = & \int_{-\infty}^\tau ds \sum_m
\d(\tau+\s-2 m l-s)f^I(s+l)\nonumber\\
& = & {1\ov 2l} \int_{-\infty}^\tau \sum_m e^{im\pi(\tau-\s-s)/l} f^I(\s+l) \ ,
\ea
we finally obtain the unitary operator
\be
U(\tau)=e^{iA(\tau)} \ ,\qq A(\tau)
=-\sqrt{\pi\ov l^3} \int_{-\infty}^{\tau+l} ds
\sum_n (-1)^n e^{-i n \pi {s\ov l}} f^I(s) a_n^I\ .
\ee
For the $S$-matrix we have
\be
S=e^{i A(\infty)}\ ,\qq A(\infty)=
-2\left(\pi\ov l\right)^{3/2} \sum_n (-1)^n \tilde f^I(n\pi/l) a_n^I \ ,
\ee
where
\be
\tilde f^I(k)= {1\ov 2\pi}
\int^{+\infty}_{-\infty} d\tau\ e^{-ik\tau} f^I(\tau)\ ,
\ee
are the Fourier components of the function $f(\tau)$.
Putting the expression for the $S$-matrix in a normal ordered form,
results into
\be
S=e^{-\d} :e^{iA(\infty)}:\ ,\qq
\d={2\pi^2\ov l^3}\sum_{n=1}^\infty n |\tilde f^I(n\pi/l)|^2\ .
\label{dpfs}
\ee
The factor $e^{-\d}$ has the interpretation of being the probability amplitude
for the string to stay in its ground state.
Moreover we may compute the expectation value of the mass square in the
remote future assuming that the string was in each ground state in the remote
past.
The Hamiltonian is simply
\be
H_{ \infty}={(A^I+B^I)^2\ov 2 l^2} + {a_0^a a_0^a\ov 2} +{\pi\ov l}
\sum_{n=1} a^i_n a^i_n\ .
\ee
Therefore the desired expectation value is
\be
M^2  =   \langle 0| S^{-1} H_{\infty} S |0 \rangle =
{(A^I+B^I)^2\ov 2 l^2} + {4\pi^4\ov l^4} \sum_{n=1}^\infty n^2
|\tilde f^I(n\pi/l)|^2  \ .
\ee
Note that in the point particle limit $l\to 0$ there should be no string
excitations at all and $\d$ should tend to zero.
Indeed, this is true for all functions
$f^I$ provided that they have Fourier coefficients that go to zero fast enough.
However, there is a correlated limit which besides sending the
length of the string to zero, it also keeps finite the mass of the string
associated with the pulse.
As a explicit illustrative example consider the profile
\be
f^I(\tau)=f_0^I {\tau_0/\pi\ov \tau_0^2+\tau^2}\ ,
\label{djh5}
\ee
with Fourier components
\be
\tilde f^I(k)={f_0^I\ov 2\pi} e^{-\tau_0|k|}\ .
\ee
Also we take $B^I=0$ so that in the far past and future the string is streched
in the same length.
Then we compute that
\be
\d = {\pi \ov 8l^3} {(f^I_0)^2\ov \sinh^2(\pi \tau_0/l)}\ .
\label{dh}
\ee
In addition, we find that the expectation value of the
mass square is
\be
M^2 = {A^I A^I\ov 2 l^2}+ {\pi^4 (f_0^I)^2\ov 4 l^4} {\cosh (\pi \tau_0/l)\ov
\sinh^3(\pi \tau_0/l)}\ .
\label{dsjh}
\ee
The first term is the usual contribution of the stretched string whereas
the second is due to the pulse.
Clearly, sending $l\to 0$ and at the same time keeping $\tau_0/l=\const$ and
$f^I_0/l^2=\const$ results from \eqn{dh} into $\d=0$ (also the $S$-matrix
becomes the identity). Then from \eqn{djh5} we see that the pulse becomes
a $\d$-function, but of vanishing strength. On the basis of these we might
have expected to obtain just the mass corresponding to the stretched string.
However, note that the second, due to the pulse, term in \eqn{dsjh}
remains finite in this correlated limit.
Therefore there is a non-trivial effect even though since $\d=0$ and $S=1$ the
probability of the string staying in its
ground state is a certainty.\footnote{A similar conclusion can be reached
with a profile similar to that in \eqn{ewjh}. However, we chose not to
present the details since the computation involves PolyGamma functions and
zeta-function regularization of infinite sums.}
We note that the limit we considered should be taken with the order we have
explained since if we simply take $S=1$ from the very beginning,
the second term in \eqn{dsjh} will not arise.
This term comes from the contribution of the whole massive tower of
string states, hence the correlated limit that we took is not really
a point particle limit.
We think that this
mechanism that gives a non-trivial contribution to the mass in
a seemingly trivial set up,
could be at work in understanding the emergence of shock waves in supergravity
solutions from a microscopic point of view as an
integrated macroscopic backreaction effect. Obviously more work in this
direction is required.

\section{Concluding remarks}
\setcounter{equation}{0}
\renewcommand{\theequation}{\thesection.\arabic{equation}}

In this paper we emphasized the possibility to promote constant moduli
parameters appearing in supergravity duals of
supersymmetric gauge theories into arbitrary
functions of the light-cone time.
We explicitly showed that, for all multicenter fundamental brane solutions of
M- and string theory, this can be done in a way
that respects not only the field equations but
also supersymmetry.
In our solutions the branes are located at centers that are functions of the
light-cone time.
Moreover, we showed that the global symmetries of the
solutions can be respected in what we called {\it shape invariant} motions.
In our construction shock wave propagation on the brane
gravitational background
arise by sudden changes in the location of the centers of the branes.
We gave explicit expressions for the supergravity backgrounds as
well as for scattering amplitudes of scalar fields
propagating in these geometries.
The most natural question that arises
is on the precise meaning of these
solutions, within the gauge/gravity correspondence,
on the gauge theory side. Since the brane centers correspond to vev's
of scalar fields we have a gauge theory with vev's that depend
on the light-cone time.
Expanding around such a vacuum gives as usual masses to the
scalars and gauge fields in the theory which now are functions of the
light-cone time. In that respect, a useful toy model to study
is that of a free scalar with a mass that depends on the light-cone time.
It turns out that the classical equations of motion and Green's functions
of the theory can be explicitly computed
for arbitrary mass profiles \cite{backsfe}. However, in making
precise contact with the computations based on supergravity solutions for
continuous brane distributions, we need to take into
account the entire mass matrix of the matrix valued scalar fields.
Moreover, we have to perform the continuous limit of the vev
distribution on the gauge theory side.
In our investigations it is important, though quite special, to give an
interpretation
on the gauge theory side of the shock wave on the maximally supersymmetric
spaces of string and M-theory of the type $AdS_p \times S^q$. Also, note the
contribution to the mass spectrum of the string modes as a reaction
to an external pulse that survived the point particle-like correlated limit
(see section 8). Perhaps this can be in the root of a mechanism
generating a shock wave
classical geometry as an integrated backreaction string effect.
We hope to report work along these directions in the future \cite{backsfe}.


\bs\bs

\centerline {\bf Acknowledgments}

\no
I would like to thank C. Bachas for collaboration on related issues.
I also acknowledge the financial support provided through the European
Community's program ``Fundamental Forces and Symmetries of the Universe''
with contract MRTN-CT-2004-005104,
the INTAS contract 03-51-6346 ``Strings, branes and higher-spin gauge
fields'', as
well as the Greek Ministry of Education programs
$\rm \P Y\Th A\G OPA\S$ with contract 89194 and
$\rm E\Pi A N$.


\vfill\eject
\appendix
\section{APPENDIX: Spin connection and Ricci tensor}
\setcounter{equation}{0}
\renewcommand{\theequation}{\thesection.\arabic{equation}}
In this appendix we compute the spin connection and the Ricci tensor
for the general metric \eqn{me1}, since these are necessary for working
out the Killing
spinor equation \eqn{Kiill} as well as the Einstein equations of motion.

\subsection{The spin connection}

Using the structure equations $de+\om\wedge e=0$ and the frame \eqn{frra}
we find for the spin connection
\ba
&& \om^{ij}=-{\a+1 \ov 2 H} \del_{[i} H dx_{j]}-{1\ov 2 H}
\del_{[i} V_{j]} du\ ,
\nonumber\\
&&  \om^{i-}={\a+1\ov 2} H^{-1/2} \dot H dx^i -\ha H^{-1/2}
\Big[\del_{[i} V_{j]} dx^j + (\del_i F -2 \dot V_i) du\Big]
\nonumber\\
&& \phantom{xxxxx}
-{\a\ov 2} H^{-3/2} \del_i H \Big[dv + V\cdot dx^i +{F\ov 2} du \Big] \ ,
\nonumber\\
&& \om^{i+}=-{\a\ov 2} H^{-3/2} \del_iH du\ ,\qq
 \om^{+-}={\a\ov 2} {\dot H\ov H} du \ ,
\label{jh6f}\\
&& \om^{\a-}={\a\ov 2} {\dot H\ov H} dx^\a \ ,
\qq \om^{\a i}= {\a\ov 2} H^{-3/2} \del_i H dx^\a\ .
\nonumber
\ea
where we have denoted tangent and target space space indices by
$i,j$ and $\a$ (not to be confused with the numerical parameter $\a$ in
\eqn{me1}).

\subsection{The Ricci tensor}

The components of the Ricci tensor are
\ba
 R_{ij} & = & a_1 {\del_i H \del_jH\ov H^2} + a_2 {\del_i \del_j H\ov H} +
\d_{ij} \Big[a_3 {\del^2 H\ov H} + a_4 {(\del H)^2\ov H^2}\Big]\ ,
\nonumber\\
&& a_1 = {1\ov 4} [ 3 (d-2) +\a (\a+4) (D-2) ]\ ,
\qq a_2= 1-\ha [d+\a (D-2)]\ ,
\nonumber\\
&& a_3 = -\ha(\a+1)\ ,
\qq a_4=-{1\ov 4} (\a+1)[ d-4+\a (D-2)]\ ,
\ea
\ba
R_{iu}  & = &
b_1 {\dot H \del_i H\ov H^2} +b_2  {\del_i \dot H\ov H}
+b_3 {1\ov  H} (\del_i \del\cdot V -\del^2 V_i)
+ b_4 {1\ov H^2} \del_{[i} V_{j]} \del_j H
\nonumber\\
&&\phantom{xxx} + b_5 {(\del H)^2\ov H^3} V_i
+ b_6 {\del^2 H\ov H^2 } V_i \ ,
\nonumber\\
&& b_1= {1\ov 4} [2 (d-1) + \a(\a+3) (D-2)]\ ,
\nonumber\\
&& b_2=-\ha[(d-1)+\a (D-2)] \ ,
\\
&& b_3=\ha \ ,\qq  b_4={1\ov 4}[d-4+\a (D-2)]\ ,
\nonumber\\
&&  b_5=-{\a\ov 4}[d-4+\a(D-2)]\ ,\qq  b_6=-{\a\ov 2}\ ,
\nonumber
\ea
\ba
R_{uu} & = & c_1 {\dot H^2\ov H^2}+c_2  {\ddot H\ov H}
+ c_3 {1\ov  H^2} (\del_{[i} V_{j]})^2
+ c_4 {1\ov H^2} \del_i H (\del_i F -2 \dot V_i)
\nonumber\\
&& \phantom{xxxxx}
+ c_5  {1\ov  H} (\del^2 F - 2 \del\cdot \dot V) + c_6 F {\del^2 H\ov H^2}
+ c_7 F {(\del H)^2\ov H^3}\ ,
\nonumber\\
&& c_1 = {1\ov 4} [d + \a (\a+2) (D-2)]\ ,\qq c_2 = -\ha [d+\a(D-2)]\ ,
\nonumber\\
&& c_3 = {1\ov 4}\ ,\qq c_4=-{1\ov 4} [d-2 + \a (D-2)]\ ,
\label{djask}\\
&& c_5 = -\ha \ , \qq c_6 = -{\a\ov 2} \ ,\qq c_7= -{\a\ov 4} [d-4+\a(D-2)]\ ,
\nonumber
\ea
\ba
R_{uv} & = & d_1 {(\del H)^2 \ov H^3} + d_2 {\del^2 H\ov H^2}\ ,
\nonumber\\
R_{\a\b} & = & \d_{\a\b}\Big[d_1 {(\del H)^2 \ov H^3} + d_2 {\del^2 H\ov H^2}
\Big]\ ,
\\
&& d_1 = -{\a\ov 4} [d-4+ \a (D-2)]\ ,\qq d_2 = -{\a \ov 2}\ .
\nonumber
\ea

\subsection{The branes}

The various coefficients above for the branes
that appear in string and M-theory are
\ba
&& {\bf M2\!-\!brane}: \qq D=11 \ , \qq d=8 \ , \qq \a=-{2\ov 3} \ ,
\nonumber\\
&& a_1=-\ha \ , \quad a_2=0 \ ,\quad a_3=-{1\ov 6} \ ,\quad a_4={1\ov 6}\ ,
\nonumber\\
&& b_1=0 \ ,\quad b_2=-\ha \ , \quad b_3=\ha \ ,\quad b_4=-\ha\ ,\quad
b_5=-{1\ov 3}\ ,
\quad b_6 = {1\ov 3}\ ,
\\
&& c_1=0 \ ,\quad c_2=-1 \ , \quad c_3={1\ov 4} \ ,\quad c_4=0 \ ,\quad
c_5=-\ha\ , \quad c_6 ={1\ov 3}\ ,\quad c_7=-{1\ov 3}\ ,
\nonumber\\
&& d_1=-{1\ov 3} \ ,\quad d_2={1\ov 3} \ .
\nonumber
\ea
\ba
&& {\bf M5\!-\!brane}: \qq D=11 \ , \qq d=5 \ , \qq \a=-{1\ov 3} \ ,
\nonumber\\
&& a_1=-\ha \ , \quad a_2=0 \ ,\quad a_3=-{1\ov 3} \ ,\quad a_4={1\ov 3}\ ,
\nonumber\\
&& b_1=0 \ ,\quad b_2=-\ha \ , \quad b_3=\ha \ ,\quad b_4=-\ha\ ,
\quad b_5=-{1\ov 6}\ , \quad b_6 ={1\ov 6}
\\
&& c_1=0 \ ,\quad c_2=-1 \ , \quad c_3={1\ov 4} \ ,\quad c_4=0\ ,
\quad c_5=-\ha\ ,
\quad c_6 ={1\ov 6}\ ,\quad c_7=-{1\ov 6}\ ,
\nonumber\\
&& d_1=-{1\ov 6} \ ,\quad d_2={1\ov 6} \ .
\nonumber
\ea
\ba
&& {\bf Dp\!-\!branes}\ (\rm string\ frame):
\qq D=10 \ , \qq d=9-p \ , \qq \a=-\ha \ ,
\nonumber\\
&& a_1={1\ov 4}(7-3 p) \ , \quad a_2=\ha(p-3) \ ,\quad a_3=-{1\ov 4} \ ,
\quad a_4={1\ov 8}(p-1) \ ,
\nonumber\\
&& b_1=\ha(3-p) \ ,\quad b_2=\ha(p-4) \ , \quad b_3=\ha \ ,\quad b_4={1\ov 4}(1-p)\ ,
\nonumber\\
&& b_5={1\ov 8}(1-p)\ , \quad b_6 ={1\ov 4}\ ,
\\
&& c_1={1\ov 4}(3-p) \ ,\quad c_2=\ha(p-5) \ , \quad c_3={1\ov 4} \ ,\quad c_4={1\ov 4}(p-3)\ ,\quad c_5=-\ha\ ,
\nonumber\\
&&  c_6 ={1\ov 4}\ ,\quad c_7={1\ov 8}(1-p)\ ,
\nonumber\\
&& d_1={1\ov 8}(1-p) \ ,\quad d_2={1\ov 4} \ .
\nonumber
\ea
\ba
&& {\bf NS1\!-\!string}\ (\rm string\ frame):
 \qq D=10 \ , \qq d=8 \ , \qq \a=-1 \ ,
\nonumber\\
&& a_1=-{3\ov 2} \ , \quad a_2=1 \ ,\qq a_3=0 \ ,\quad a_4=0\ ,
\nonumber\\
&& b_1=-\ha \ ,\quad b_2=\ha \ ,\quad b_3=\ha \ ,\quad b_4=-1\ ,\quad
b_5=-1\ ,\quad b_6 =\ha
\\
&& c_1=0 \ ,\quad c_2=0 \ , \quad c_3={1\ov 4} \ ,\quad c_4=\ha\ ,\quad c_5=-\ha\ , \quad c_6 =\ha\ ,\quad c_7=-1\ ,
\nonumber\\
&& d_1=-1 \ ,\quad d_2=\ha \ .
\nonumber
\ea
\ba
&& {\bf NS5\!-\!brane}\ (\rm string\ frame):
\qq D=10 \ , \qq d=4 \ , \qq \a=0 \ ,
\nonumber\\
&& a_1={3\ov 2} \ , \quad a_2=-1 \ ,\quad a_3=-\ha \ ,\quad a_4=0 \ ,
\nonumber\\
&& b_1={3\ov 2} \ ,\quad b_2=-{3\ov 2} \ , \quad b_3=\ha \ ,\quad b_4=0\ ,\quad
 b_5=0\ , \quad b_6 =0 \ ,
\\
&& c_1=1 \ ,\quad c_2=-2 \ , \quad c_3={1\ov 4} \ ,\quad c_4=-\ha\ ,\quad
 c_5=-\ha\ , \quad c_6 =0\ ,\quad c_7=0\ ,
\nonumber\\
&& d_1= 0\ ,\quad d_2=0 \ .
\nonumber
\ea

\vfill\eject

\end{document}